\documentclass{aa}  

\usepackage{graphicx}
\usepackage{txfonts}
\usepackage{lipsum}
\usepackage{subcaption}         
\usepackage{lscape}             
\usepackage{placeins}           
                                
\usepackage{natbib}
\bibpunct{(}{)}{;}{a}{}{,} 
\usepackage{hyperref}
\usepackage{multicol}

\begin{document}

   \title{A new framework for treating the ortho-to-para ratio of H$_2$ and H$_3^+$ isotopologues in astrochemical models}

   \titlerunning{A new framework for treating the ortho-to-para ratio of H$_2$ and H$_3^+$ isotopologues}


   \author{Y. Komichi\inst{1}
        \and K. Furuya\inst{2}
        }

   \institute{Department of Astronomy, the University of Tokyo, 7-3-1 Hongo, Bunkyo-ku, Tokyo 113-0033, Japan\\
             \email{komichi-yuto832@g.ecc.u-tokyo.ac.jp}
            \and  RIKEN Pioneering Research Institute, 2-1 Hirosawa, Wako-shi, Saitama 351-0198, Japan\\ }

   \date{Received September 30, 20XX}

 
  \abstract
   {The time evolution of the ortho-to-para ratios of H$_2$ and H$_3^+$ isotopologues in star-forming regions has been investigated using astrochemical models with chemical networks in which reactions are distinguished according to their nuclear spin isomers. This conventional approach requires the original chemical network to be modified and extended to explicitly treat nuclear spin isomers.}
   {We develop a new framework to study the time evolution of the ortho-to-para ratio of H$_2$ and H$_3^+$ isotopologues. This framework does not require reactions to be distinguished according to their nuclear spin isomers and can be applied directly to the original chemical network without modification.} 
   {We derive time evolution equations for ortho-to-para ratios that can be solved simultaneously with any chemical network in which chemical reactions are not explicitly separated according to the nuclear spin isomers of H$_2$ and H$_3^+$ isotopologues.}
   {We conducted test calculations using our framework under static and dynamical physical conditions and compared the results with those obtained by the conventional method. We find that the ortho-to-para ratios calculated with the two methods show the typical error of up to 26\%, which supports the validity of the framework over the tested parameter range. We also confirm that the total abundances summed over all nuclear spin states for each molecule are almost identical between the two methods.}
   {The new framework allows us to include nuclear spin chemistry of H$_2$ and H$_3^+$ isotopologues in any astrochemical models without modifying the original chemical network. In addition to the conventional approach, our framework provides an alternative approach for studying the nuclear spin chemistry in star-forming regions.}


   \maketitle
   \nolinenumbers

\section{Introduction} \label{sec:intro}

Molecular hydrogen, which is the primary constituent of molecular clouds, exists in two nuclear spin isomers: ortho-H$_2$ (o-H$_2$), with parallel nuclear spins, and para-H$_2$ (p-H$_2$), with antiparallel spins. Since spontaneous transitions between ortho and para-H$_2$ are forbidden, the conversion of the two states proceeds via chemical reactions or nuclear spin conversions on solid surfaces \citep[e.g., ][]{Fukutani_2013}. Thus, they effectively behave as distinct molecules.

The energy difference between their ground states in the gas phase is approximately 170\,K with o-H$_2$ being the higher energy state. The energy difference is much larger than the typical temperature of molecular clouds ($\sim 10\,\mathrm{K}$). Therefore, the abundance ratio of o-H$_2$ to p-H$_2$ (hereafter the H$_2$ ortho-to-para ratio, or the H$_2$ OPR) affects both the physical and chemical properties of molecular clouds. Physically, the equation of state of molecular gas \citep[e.g.][]{Decampli_1978,Boley_2007,Vaytet_2014} and the cooling rate by rotational and vibrational lines from H$_2$ \citep[e.g.][]{Glover_2008} depend on the H$_2$ OPR. In terms of chemistry, the efficiency of deuterium fractionation depends on the H$_2$ OPR \citep[e.g.,][]{Pagani_1992}. For example, the most important isotope exchange reaction for triggering the deuterium fractionation is
\begin{equation}
    \text{H}_3^+ + \text{HD} \rightleftharpoons \text{H}_2\text{D}^+ + \text{H}_2 + \Delta E,
\end{equation}
where $\Delta E$ is the heat of the reaction, which depends on the nuclear spin configurations of H$_3^+$, H$_2$D$^+$, and H$_2$. The backward reaction is endothermic, and does not proceed efficiently in molecular cloud conditions when the H$_2$ OPR is small ($\lesssim 10^{-3}$). On the other hand, if the H$_2$ OPR is high ($\gtrsim 10^{-3}$) and thus o-H$_2$ is sufficiently abundant, the backward reaction becomes efficient, slowing down the deuterium fractionation.

Direct observations of H$_2$ are limited to hot ($\gtrsim 10^2\,\mathrm{K}$) regions such as photodissociation regions (PDRs) and shocked gas where H$_2$ is rotationally or vibrationally excited. Observations of H$_2$ in PDRs indicate that the H$_2$ OPR is close to the statistical value of three or somewhat lower \citep[e.g.,][]{Sternberg_1999,Zannese_2025}. On the other hand, observations of H$_2$ in cold molecular clouds are challenging, and the H$_2$ OPR has been estimated from the observations of deuterated species and astrochemical modeling \citep[e.g.,][]{Maret_2007,Pagani_2009,Pagani_2013,Brunken_2014}. 

\begin{figure*}[h!]
    \centering
    \includegraphics[width=\hsize]{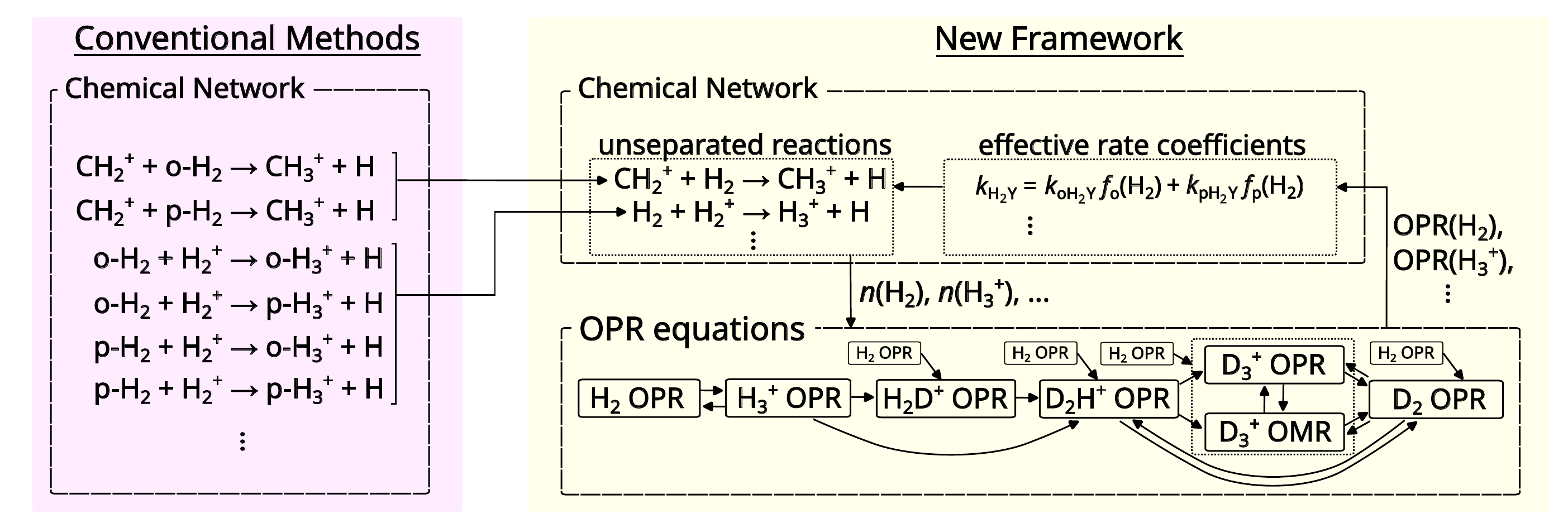}
    \caption{The concept of our new framework compared with the conventional methods to calculate the nuclear spin chemistry of H$_2$ and H$_3^+$. The capital Y that appears in $k_\mathrm{H_2Y}$ represents a certain molecule Y (see also equation (\ref{eq:k_XY})).}
    \label{fig:OPRconcept}
\end{figure*}

There are many previous studies that have performed time-dependent astrochemical network calculations considering the nuclear spin chemistry of H$_2$ and H$_3^+$, together with deuterium fractionation chemistry \citep[e.g.][]{Walmsley_2004,Flower_2006,Sipila_2013,Furuya_2015,Majumdar_2017,Aikawa_2018}. In those previous studies, different nuclear spin isomers are treated as distinct species in the chemical network, which significantly increases the number of reactions. They rewrite the reactions in the original chemical network, in which nuclear spin isomers are not considered, by explicitly separating H$_2$, H$_3^+$, and their isotopologues into their respective nuclear spin isomers (Figure \ref{fig:OPRconcept}). They also add spin-exchange reactions to the chemical network. This conventional method is straightforward in theory. However, since they need to split all reactions containing H$_2$, H$_3^+$, and their isotopologues, such a network model requires a vast number of chemical reactions. This increases the complexity of the network model, particularly when the original chemical network already contains a lot of reactions. In addition, reactions must be separated whenever the chemical networks are modified or updated. The modification of existing chemical networks to explicitly treat nuclear-spin states can be laborious and time-consuming. This motivates the development of a framework that can account for nuclear spin chemistry while retaining the structure of existing networks.

As an alternative approach, we present a new framework for treating the H$_2$ OPR in any astrochemical models without explicitly including nuclear spin isomers in chemical networks. We propose to solve the time evolution equation for the H$_2$ OPR (hereafter, the H$_2$ OPR equation), together with gas (and ice) chemical network that does not explicitly distinguish the spin isomers (Figure \ref{fig:OPRconcept}). This allows us to obtain the self-consistent solution of the H$_2$ OPR and the abundances of chemical species in the network. In addition to the H$_2$ OPR, we also present a simple method to estimate the OPRs of D$_2$ and H$_3^+$ isotopologues, which play a key role in calculating the deuterium fractionation chemistry.

The rest of this paper is organized as follows. The construction of the OPR equation for H$_2$, H$_3^+$, and their isotopologues is described in Section \ref{sec:H2OPReq}. The setup and results of the test calculations are shown in Section \ref{sec:testcalc}. We discuss the limitations and advantages of our framework in Section \ref{sec:discussion}. Finally, we summarize the present work in Section \ref{sec:summary}.

\section{Formulation of new framework}\label{sec:H2OPReq}

\subsection{Concept of new framework and OPR equations}\label{sec:basiceq}

As described in the previous section, the conventional methods solve the chemical network in which different nuclear spin isomers are treated as distinct species, and the reactions are explicitly separated by the nuclear spin isomers. Instead, our new framework self-consistently solves both the OPR equations and the chemical network in which reactions are not explicitly separated by nuclear spin isomers (Figure \ref{fig:OPRconcept}). We note that the details of the chemical reactions are described in Tables \ref{table:OPRreaction} to \ref{table:OPRreaction_D2}.

The general rate equations for ortho and para species (o-X and p-X, respectively) are given as follows:
\begin{equation}\label{eq:orthoX}
    \begin{split}
        \frac{\mathrm{d}}{\mathrm{d} t} n(\text{o-X})=&F(\text{o-X})+K_\mathrm{p\rightarrow o}n(\text{p-X})\\
        &-K_\mathrm{o\rightarrow p}n(\text{o-X}) - D(\text{o-X})n(\text{o-X}),
    \end{split}
\end{equation}
and
\begin{equation}
    \begin{split}
        \frac{\mathrm{d}}{\mathrm{d} t} n(\text{p-X})=&F(\text{p-X})+K_\mathrm{o\rightarrow p}n(\text{o-X})\\
        &-K_\mathrm{p\rightarrow o}n(\text{p-X}) - D(\text{p-X})n(\text{p-X}),
    \end{split}
\end{equation}
where $n(\mathrm{X})$ is the number density of X, $F(\text{X})$ is the total formation rate of X, $K_\mathrm{o\rightarrow p}$ ($K_\mathrm{p\rightarrow o}$) is the conversion rate coefficient from ortho (para) to para (ortho), and $D(\text{X})$ is the total destruction rate coefficient of X. We note that the conversion rates ($K_\mathrm{o\rightarrow p}$ and $K_\mathrm{p\rightarrow o}$) are not included in the total formation rate, $F(\text{X})$. The time evolution equation for the ortho-to-para ratio of X, $\text{OPR(X)}=n(\text{o-X})/n(\text{p-X})$, can be written as
\begin{equation}\label{eq:OPRX}
     \begin{split}
         \frac{\mathrm{d}}{\mathrm{d} t} \text{OPR(X)}=&\frac{1}{n(\text{p-X})}\frac{\mathrm{d}}{\mathrm{d} t} n(\text{o-X})-\frac{n(\text{o-X})}{(n(\text{p-X}))^2}\frac{\mathrm{d}}{\mathrm{d} t} n(\text{p-X})\\
         =&-\left[K_\mathrm{o\rightarrow p}+\frac{F(\text{p-X})}{n(\text{X})}\right][\text{OPR(X)}]^2 \\
         &-\left[K_\mathrm{o\rightarrow p}-K_\mathrm{p\rightarrow o}-\frac{F(\text{o-X})}{n(\text{X})}\right. \\
         &\ \ \ \left.+\frac{F(\text{p-X})}{n(\text{X})}+ D(\text{o-X})-D(\text{p-X})\right]\text{OPR(X)} \\
         &+K_\mathrm{p\rightarrow o}+\frac{F(\text{o-X})}{n(\text{X})},
     \end{split}
\end{equation}
where $n(\text{X})=n(\text{o-X})+n(\text{p-X})$. 
The equations depend on the formation rate of each molecule. Therefore, we solve the equations by referring to the formation rate calculated by the network. 

Some chemical reactions (e.g., H$_2$D$^+$ + H$_2$ $\rightarrow$ H$_3^+$ + HD and N$^+$ + H$_2$ $\rightarrow$ NH$^+$ + H) have different rate coefficients depending on the nuclear spin configurations. We treat such reactions as exceptions. For the exceptional reactions, we calculate the effective rate coefficient weighted by the fraction of the abundance of each nuclear spin isomer in the chemical network calculations, using the OPRs obtained from the OPR equations. If a molecule X has ortho and para nuclear spin configurations, the effective rate coefficient of the reaction between X and Y, $k_\mathrm{XY}$, is given as
\begin{equation}\label{eq:k_XY_derive}
    \begin{split}
        &k_\mathrm{XY}n(\text{X})n(\text{Y}) \\
        =&k_\mathrm{oXY}n(\text{o-X})n(\text{Y})+k_\mathrm{pXY}n(\text{p-X})n(\text{Y}) \\
        =& \left[k_\mathrm{oXY}\frac{n(\text{o-X})}{n(\text{o-X})+n(\text{p-X})}+k_\mathrm{pXY}\frac{n(\text{p-X})}{n(\text{o-X})+n(\text{p-X})}\right]n(\text{X})n(\text{Y})\\
        =&\left[k_\mathrm{oXY}\frac{\text{OPR}(\text{X})}{1+\text{OPR}(\text{X})}n(\text{X})n(\text{Y})+k_\mathrm{pXY}\frac{1}{1+\text{OPR}(\text{X})}\right]n(\text{X})n(\text{Y})\\
        =&\left[k_\mathrm{oXY}f_\mathrm{o}(\text{X})+k_\mathrm{pXY}f_\mathrm{p}(\text{X})\right]n(\text{X})n(\text{Y}),
    \end{split}
\end{equation}
therefore, 
\begin{equation}\label{eq:k_XY}
    \begin{split}
        k_\mathrm{XY}=k_\mathrm{oXY}f_\mathrm{o}(\text{X})+k_\mathrm{pXY}f_\mathrm{p}(\text{X}),
    \end{split}
\end{equation}
where $k_\mathrm{oXY}$ $(k_\mathrm{pXY})$ is the rate coefficient of the reaction between o-X (p-X) and Y, and $f_\mathrm{o}(\text{X})=\text{OPR}(\text{X})/[1+\text{OPR}(\text{X})]$ $(f_\mathrm{o}(\text{X})=\text{OPR}(\text{X})/[1+\text{OPR}(\text{X})])$ is the fraction of o(p)-X abundance with respect to the total X abundance.

We note that our approach is similar to some previous studies (e.g., \citet{Cleeves_2014,Lee_2015}), which also include the H$_2$ OPR in the chemical network where chemical reactions are not split by the nuclear spin isomers. However, they assume the H$_2$ OPR and H$_3^+$ OPR to be at their thermal equilibrium values at given temperatures, while we solve the time evolution of the H$_2$ OPR and H$_3^+$ OPR self-consistently with the chemical network. In addition, although the description of the effective rate coefficients weighted by OPR given as equation (\ref{eq:k_XY}) is mathematically valid, we need to solve both the fractional abundances of chemical species and their OPRs at the same time, since both of them vary with time simultaneously.

In the rest of this section, we discuss more details on the OPR equations for H$_2$ (Section \ref{sec:H2OPR}), H$_3^+$ (Section \ref{sec:H3+OPR}), H$_2$D$^+$ and D$_2$H$^+$ (Section \ref{sec:H2D+OPR}), D$_3^+$ (Section \ref{sec:D3+OPR}), and D$_2$ (Section \ref{sec:D2OPR}).

\subsection{H$_2$}\label{sec:H2OPR}

In the case of H$_2$, both H$_2$ in the gas phase and on grain surfaces should be considered, since the H$_2$ formation on grain surfaces is essential. Following Appendix A of \citet{Furuya_2015}, the rate equations for o-H$_2$ in the gas phase and on grain surfaces are given by
\begin{equation}\label{eq:orthoH2}
    \begin{split}
        \frac{\mathrm{d}}{\mathrm{d} t} n(\text{o-H}_2)=&F(\text{o-H}_2)+K_\mathrm{p\rightarrow o}n(\text{p-H}_2)\\
        &-K_\mathrm{o\rightarrow p}n(\text{o-H}_2) - D(\text{o-H}_2)n(\text{o-H}_2),
    \end{split}
\end{equation}
and 
\begin{equation}\label{eq:orthoJH2}
    \begin{split}
        \frac{\mathrm{d}}{\mathrm{d} t} n(\text{Jo-H}_2)=&F(\text{Jo-H}_2) - D(\text{Jo-H}_2)n(\text{Jo-H}_2),
    \end{split}
\end{equation}
where species prefixed with "J" denote molecular species on grain surfaces. The total formation and destruction rate of o-H$_2$ and Jo-H$_2$ are given by
\begin{equation}\label{eq:FoH2}
    F(\text{o-H}_2)=k_\mathrm{o}^\mathrm{des}n(\text{Jo-H}_2)+\mathcal{F}(\text{o-H}_2),
\end{equation}
\begin{equation}\label{eq:DoH2}
    D(\text{o-H}_2)=k_\mathrm{o}^\mathrm{ads}+\mathcal{D}(\text{o-H}_2),
\end{equation}
\begin{equation}\label{eq:FJoH2}
    F(\text{Jo-H}_2)=k_\mathrm{o}^\mathrm{ads}n(\text{Jo-H}_2)+\mathcal{F}(\text{Jo-H}_2),
\end{equation}
and
\begin{equation}
    D(\text{Jo-H}_2)=k_\mathrm{o}^\mathrm{des},
\end{equation}
where $k_\mathrm{o}^\mathrm{des}$ and $k_\mathrm{o}^\mathrm{ads}$ are the desorption and adsorption rate coefficients of o-H$_2$, respectively, $\mathcal{F}(\text{X})$ and $\mathcal{D}(\text{X})$ are the total formation and destruction rates, respectively, of X via gas-phase reactions only (e.g., ion-neutral reactions and photodissociation), and $\mathcal{F}(\text{JX})$ is the total formation rate of JX via grain surface reactions only. The summation of equations (\ref{eq:orthoH2}) and (\ref{eq:orthoJH2}) results in
\begin{equation}\label{eq:totoH2}
    \begin{split}
        \frac{\mathrm{d}}{\mathrm{d} t} [n(\text{o-H}_2)+n(\text{Jo-H}_2)]=&\mathcal{F}^\mathrm{tot}(\text{o-H}_2)+K_\mathrm{p\rightarrow o}n(\text{p-H}_2)\\
        &-K_\mathrm{o\rightarrow p}n(\text{o-H}_2) - \mathcal{D}(\text{o-H}_2)n(\text{o-H}_2),
    \end{split}
\end{equation}
where $\mathcal{F}^\mathrm{tot}(\text{X})=\mathcal{F}(\text{X})+\mathcal{F}(\text{JX})$ is the total formation rate of X and JX via gas-phase and grain surface reactions. Since H$_2$ on grain surfaces is much less abundant than H$_2$ in the gas phase, we assume that $n(\text{o-H}_2)+n(\text{Jo-H}_2)\simeq n(\text{o-H}_2)$. Then equation (\ref{eq:totoH2}) is rewritten as
\begin{equation}\label{eq:totoH2_}
    \begin{split}
        \frac{\mathrm{d}}{\mathrm{d} t} n(\text{o-H}_2)=&\mathcal{F}^\mathrm{tot}(\text{o-H}_2)+K_\mathrm{p\rightarrow o}n(\text{p-H}_2)\\
        &-K_\mathrm{o\rightarrow p}n(\text{o-H}_2) - \mathcal{D}(\text{o-H}_2)n(\text{o-H}_2).
    \end{split}
\end{equation}
The rate equation for p-H$_2$ is given in the same form as equation (\ref{eq:totoH2_}). The time evolution equation for H$_2$ OPR can be written as
\begin{equation}\label{eq:OPRH2}
     \begin{split}
         \frac{\mathrm{d}}{\mathrm{d} t} \text{OPR(H}_2\text{)}=&-\left[K_\mathrm{o\rightarrow p}+\frac{\mathcal{F}^\mathrm{tot}(\text{p-H}_2\text{})}{n(\text{H}_2\text{})}\right][\text{OPR(H}_2\text{)}]^2 \\
         &-\left[K_\mathrm{o\rightarrow p}-K_\mathrm{p\rightarrow o}-\frac{\mathcal{F}^\mathrm{tot}(\text{o-H}_2\text{})}{n(\text{H}_2\text{})}\right. \\
         &\ \ \ \left.+\frac{\mathcal{F}^\mathrm{tot}(\text{p-H}_2\text{})}{n(\text{H}_2\text{})}+ \mathcal{D}(\text{o-H}_2\text{})-\mathcal{D}(\text{p-H}_2\text{})\right]\text{OPR(H}_2\text{)} \\
         &+K_\mathrm{p\rightarrow o}+\frac{\mathcal{F}^\mathrm{tot}(\text{o-H}_2\text{})}{n(\text{H}_2\text{})}.
     \end{split}
\end{equation}

The ortho-para conversion of H$_2$ mainly proceeds via gas-phase reactions with H$^+$ and H$_3^+$, as well as nuclear spin conversion (NSC) on grain surfaces \citep[e.g.,][]{Gerlich_1990,Honvault_2011,Furuya_2019}. Table \ref{table:OPRreaction} summarizes the exchange reactions considered in this work. When neglecting other minor exchange reactions (e.g., H$_2$ + H$_2$D$^+$ and H$_2$ + D$_2$H$^+$), the conversion rate from the ortho to the para state can be approximated as follows:
\begin{equation}\label{eq:OPRH2_conversionrate}
    K_\mathrm{o\rightarrow p}\simeq k_\mathrm{o\rightarrow p}^\mathrm{H^+}n(\text{H}^+)+k_\mathrm{o\rightarrow p}^\mathrm{H_3^+}n(\text{H}_3^+)+K_\mathrm{o\rightarrow p}^\mathrm{surf},
\end{equation}
where $k_\mathrm{o\rightarrow p}^\mathrm{H^+}$ is the rate coefficient for the gas phase reaction between o-H$_2$ and H$^+$:
\begin{equation}\label{eq:opH2&H+}
    \text{o-H}_2 + \text{H}^+ \rightarrow \text{p-H}_2 + \text{H}^+.
\end{equation}
$k_\mathrm{o\rightarrow p}^\mathrm{H_3^+}$ is the effective rate coefficient (see equation (\ref{eq:k_XY})) for reactions between o-H$_2$ and o- or p-H$_3^+$ weighted by the fraction of o-H$_3^+$ ($\text{OPR}(\text{H}_3^+)/(1+\text{OPR}(\text{H}_3^+))$) or that of p-H$_3^+$ ($1/(1+\text{OPR}(\text{H}_3^+))$):
\begin{equation}
    \begin{split}
        k_\mathrm{o\rightarrow p}^\mathrm{H_3^+}=&k_\mathrm{ooH_3^+}\frac{\text{OPR}(\text{H}_3^+)}{1+\text{OPR}(\text{H}_3^+)}+k_\mathrm{opH_3^+}\frac{1}{1+\text{OPR}(\text{H}_3^+)}\\
        =&k_\mathrm{ooH_3^+}f_\mathrm{o}(\text{H}_3^+)+k_\mathrm{opH_3^+}f_\mathrm{p}(\text{H}_3^+),
    \end{split}
\end{equation}
and $K_\mathrm{o\rightarrow p}^\mathrm{surf}$ represents the rate coefficient of NSC on grain surfaces, which is derived in \citet{Furuya_2019,Furuya_2026}. $k_\mathrm{ooH_3^+}$ and $k_\mathrm{opH_3^+}$ are the rate coefficients of the following reactions:
\begin{equation}\label{eq:oH2andoH3+}
    \text{o-H}_2 + \text{o-H}_3^+ \rightarrow \text{p-H}_2 + \text{o/p-H}_3^+,
\end{equation}
and
\begin{equation}\label{eq:oH2andpH3+}
    \text{o-H}_2 + \text{p-H}_3^+ \rightarrow \text{p-H}_2 + \text{o/p-H}_3^+,
\end{equation}
respectively. $K_\mathrm{p\rightarrow o}$ is calculated in a similar way. The way of evaluating $\text{OPR}(\text{H}_3^+)$ is described in Section \ref{sec:H3+OPR}.

When H$_2$ forms on grain surfaces, the H$_2$ OPR is the statistical value of three referring to the laboratory experiments \citep{Watanabe_2010,Tsuge_2019,Tsuge_2021}. The gas-phase chemical networks that include nuclear spin isomers in the literature \citep[e.g.,][]{Majumdar_2017} are constructed with Oka's method \citep{Oka_2004}, in which the branching ratios are obtained by angular momentum algebra. Referring to such chemical networks, the H$_2$ OPR for almost all gas-phase reactions forming H$_2$ is assumed to be three. However, there are some exceptional reactions whose rate coefficients depend on OPRs of H$_3^+$, its isotopologues, and D$_2$. For those reactions, we calculate the H$_2$ formation rates by weighing the reaction rates using the OPRs.
Therefore, the total formation rates of ortho and para-H$_2$ are given by
\begin{equation}
    \begin{split}
        \mathcal{F}^\mathrm{tot}(\text{o-H}_2)=&\mathcal{F}(\text{o-H}_2)+\mathcal{F}(\text{Jo-H}_2) \\
        =&\frac{3}{4}\mathcal{F}^\text{st}(\text{H}_2)+\mathcal{F}^\text{opr}(\text{o-H}_2)+\frac{3}{4}\mathcal{F}(\text{JH}_2),
    \end{split}
\end{equation}
and
\begin{equation}
    \begin{split}
        \mathcal{F}^\mathrm{tot}(\text{p-H}_2)=&\mathcal{F}(\text{p-H}_2)+\mathcal{F}(\text{Jp-H}_2) \\
        =&\frac{1}{4}\mathcal{F}^\text{st}(\text{H}_2)+\mathcal{F}^\text{opr}(\text{p-H}_2)+\frac{1}{4}\mathcal{F}(\text{JH}_2),
    \end{split}
\end{equation}
where $\mathcal{F}^\text{st}(\text{X})$ and $\mathcal{F}^\text{opr}(\text{X})$ are the total H$_2$ formation rates whose H$_2$ OPR is the statistical value and depends on the OPRs of other molecules, respectively. The destruction rates of ortho and para-H$_2$ in the gas phase are assumed to be identical for almost all reactions in the chemical networks described above (i.e., $\mathcal{D}(\text{o-H}_2)=\mathcal{D}(\text{p-H}_2)$). Then they cancel out in equation (\ref{eq:OPRH2}). However, similar to the formation rates, we include the destruction rates of the exceptional reactions whose rate coefficients depend on the OPRs of H$_3^+$ isotopologues and HD.

\subsection{H$_3^+$}\label{sec:H3+OPR}

\begin{table}
\caption{Chemical reactions explicitly considered in OPR equations of H$_2$}             
\label{table:OPRreaction}  
\centering                       
\begin{tabular}{l|l}    
\hline
\multicolumn{2}{c}{H$_2$ OPR (depending on H$_3^+$ OPR)} \\         
\hline                      
    Exchange    & H$^+$ + o-H$_2$ $\rightarrow$ H$^+$ + p-H$_2$ \\
                & H$^+$ + p-H$_2$ $\rightarrow$ H$^+$ + o-H$_2$ \\
                & o-H$_3^+$ + o-H$_2$ $\rightarrow$ o/p-H$_3^+$ + p-H$_2$ \\
                & p-H$_3^+$ + o-H$_2$ $\rightarrow$ o/p-H$_3^+$ + p-H$_2$ \\
                & o-H$_3^+$ + p-H$_2$ $\rightarrow$ o/p-H$_3^+$ + o-H$_2$ \\
                & p-H$_3^+$ + p-H$_2$ $\rightarrow$ o/p-H$_3^+$ + o-H$_2$ \\
\hline
    spin conversion    & o-H$_2$ $\rightarrow$ p-H$_2$  \\
    on grain surface & p-H$_2$ $\rightarrow$ o-H$_2$  \\
\hline
\end{tabular}
\tablefoot{The "Dependence" label in each subheader indicates the OPRs on which the listed reactions depend. As for H$_2$ OPR, we consider various formation reactions of H$_2$ which are not listed here (see Section \ref{sec:H2OPR}). The references can be found in Section \ref{sec:setting}.}
\end{table}

\begin{table}
\caption{Similar to Table\ref{table:OPRreaction}, but for D$_3^+$ OPR}              
\label{table:OPRreaction_D3pOPR}   
\centering                      

\begin{tabular}{l|l}     

\hline
\multicolumn{2}{c}{D$_3^+$ OPR (depending on H$_2$, D$_2$, and D$_2$H$^+$ OPR, and D$_3^+$ OMR)} \\       
\hline
    Formation   & o-D$_2$H$^+$ + HD $\rightarrow$ o-D$_3^+$ + HD \\
                & o-D$_2$H$^+$ + HD $\rightarrow$ p-D$_3^+$ + HD \\
                & p-D$_2$H$^+$ + HD $\rightarrow$ o-D$_3^+$ + HD \\
                & p-D$_2$H$^+$ + HD $\rightarrow$ p-D$_3^+$ + HD \\
                & m-D$_3^+$ + HD $\rightarrow$ o-D$_3^+$ + HD \\
                & m-D$_3^+$ + o-D$_2$ $\rightarrow$ o-D$_3^+$ + o/p-D$_2$ \\
                & m-D$_3^+$ + o-D$_2$ $\rightarrow$ p-D$_3^+$ + o/p-D$_2$ \\
                & m-D$_3^+$ + p-D$_2$ $\rightarrow$ o-D$_3^+$ + o/p-D$_2$ \\
                & m-D$_3^+$ + p-D$_2$ $\rightarrow$ p-D$_3^+$ + o/p-D$_2$ \\
\hline                     
    Exchange    & o-D$_3^+$ + HD $\rightarrow$ p-D$_3^+$ + HD \\
                & p-D$_3^+$ + HD $\rightarrow$ o-D$_3^+$ + HD \\
                & o-D$_3^+$ + o-D$_2$ $\rightarrow$ p-D$_3^+$ + o/p-D$_2$ \\
                & o-D$_3^+$ + p-D$_2$ $\rightarrow$ p-D$_3^+$ + o/p-D$_2$ \\
                & p-D$_3^+$ + o-D$_2$ $\rightarrow$ o-D$_3^+$ + o/p-D$_2$ \\
                & p-D$_3^+$ + p-D$_2$ $\rightarrow$ o-D$_3^+$ + o/p-D$_2$ \\
\hline                    
    Destruction & o-D$_3^+$ + o-H$_2$ $\rightarrow$ o/p-D$_2$H$^+$ + HD \\
                & o-D$_3^+$ + p-H$_2$ $\rightarrow$ o/p-D$_2$H$^+$ + HD \\
                & p-D$_3^+$ + o-H$_2$ $\rightarrow$ o/p-D$_2$H$^+$ + HD \\
                & p-D$_3^+$ + p-H$_2$ $\rightarrow$ o/p-D$_2$H$^+$ + HD \\
                & o-D$_3^+$ + HD $\rightarrow$ m-D$_3^+$ + HD \\
\hline                  
\end{tabular}
\end{table}

In the case of H$_3^+$, ortho-para conversion proceeds via gas-phase reactions with H$_2$ \citep[e.g.,][]{Gerlich_1990,Hugo_2009}. The conversion rate from ortho to para state is given by
\begin{equation}
    \begin{split}
        K_\mathrm{o\rightarrow p}&= k_\mathrm{ooH_2}n(\text{o-H}_2)+k_\mathrm{opH_2}n(\text{p-H}_2) \\
        &=\left[k_\mathrm{ooH_2}\frac{\text{OPR}(\text{H}_2)}{1+\text{OPR}(\text{H}_2)}+k_\mathrm{opH_2}\frac{1}{1+\text{OPR}(\text{H}_2)}\right]n(\text{H}_2) \\
        &=\left[k_\mathrm{ooH_2}f_\mathrm{o}(\text{H}_2)+k_\mathrm{opH_2}f_\mathrm{p}(\text{H}_2)\right]n(\text{H}_2),
    \end{split}
\end{equation}
where $k_\mathrm{ooH_2}$ and $k_\mathrm{opH_2}$ are the rate coefficients of the following reactions:
\begin{equation}
    \text{o-H}_2 + \text{o-H}_3^+ \rightarrow \text{o/p-H}_2 + \text{p-H}_3^+,
\end{equation}
and
\begin{equation}
    \text{p-H}_2 + \text{o-H}_3^+ \rightarrow \text{o/p-H}_2 + \text{p-H}_3^+.
\end{equation}

H$_3^+$ is primarily formed via the gas-phase reaction between H$_2$ and H$_2^+$. Therefore, the formation rate of H$_3^+$ is given by
\begin{equation}
    F(\text{H}_3^+)=k_\mathrm{H_2^++H_2}n(\text{H}_2^+)n(\text{H}_2),
\end{equation}
where $k_\mathrm{H_2^++H_2}$ is the rate coefficient for the reaction between H$_2^+$ and H$_2$.

H$_3^+$ is destroyed through various reactions. As in the case of H$_2$, the rate coefficients for the destruction reactions of o- and p-H$_3^+$ are assumed to be identical for most reactions in the chemical networks used in the conventional methods. Such reactions do not contribute to the H$_3^+$ OPR because they cancel out in equation (\ref{eq:OPRX}). The exceptions are, for example, the reactions with e$^-$ or with HD. In this work, we set $D(\text{o-H}_3^+)$ as follows:
\begin{equation}
    D(\text{o-H}_3^+)=k_{\text{o-H}_3^++\text{e}^-}n(\mathrm{e^-})+k_{\text{o-H}_3^++\text{HD}}n(\mathrm{HD}),
\end{equation}
and an analogous expression is adopted for $D(\text{p-H}_3^+)$. Here, $k_{\text{o-H}_3^++\text{e}^-}$ and $k_{\text{o-H}_3^++\text{HD}}$ are the total rate coefficients for the reactions, $\text{o-H}_3^+ + \text{e}^-$ and $\text{o-H}_3^+ + \text{HD}$, respectively, including all branches. The included reactions are summarized in Table \ref{table:OPRreaction_H3p}.

There are two possible approaches to calculate the time evolution of the H$_3^+$ OPR. One is to solve the time evolution equation for the H$_3^+$ OPR (H$_3^+$ OPR equation), coupled with the H$_2$ OPR equation. The other is to derive the quasi-steady-state solution of the H$_3^+$ OPR equation at a given time. Unlike the H$_2$ OPR, the H$_3^+$ OPR is determined via ion-molecule reactions and the electron recombination reactions, which are much faster than the timescale of the evolution of H$_2$ OPR. Therefore, the quasi-steady-state assumption for H$_3^+$ OPR at each timestep would be valid in the typical dense cloud conditions (see Section \ref{sec:limit_of_framework}). In this work, we take the second approach. By setting the time derivative term in the H$_3^+$ OPR equation to zero (i.e., $\mathrm{d}\text{OPR(H}_3^+{)}/\mathrm{d}t=0$), the quasi-steady-state H$_3^+$ OPR is obtained by solving a quadratic equation. We note that the quasi-steady-state H$_3^+$ OPR still evolves with time because the H$_2$ OPR and other terms appearing in the H$_3^+$ OPR equation vary with time.

\subsection{H$_2$D$^+$ and D$_2$H$^+$}\label{sec:H2D+OPR}

Similar to H$_3^+$, we calculate the H$_2$D$^+$ and D$_2$H$^+$ OPRs under the quasi-steady-state assumption. For H$_2$D$^+$ and D$_2$H$^+$, we include the formation, conversion, and destruction reactions listed in Table \ref{table:OPRreaction_H2Dp} and \ref{table:OPRreaction_D2Hp}.

D$_2$H$^+$ is also formed through reactions between H$_3^+$ and D$_2$. This reaction becomes important when the reaction H$_3^+$ + D$_2$ $\rightarrow$ D$_2$H$^+$ + H$_2$ more contributes to D$_2$H$^+$ formation than the reaction H$_2$D$^+$ + HD $\rightarrow$ D$_2$H$^+$ + H$_2$, i.e., $n(\text{H}_3^+)n(\text{D}_2)\gtrsim n(\text{H}_2\text{D}^+)n(\text{HD})$. Since D$_2$ also has spin isomers and D$_2$ OPR depends on D$_2$H$^+$ OPR through D$_3^+$ OPR, for example, we need iteration to obtain the D$_2$H$^+$ OPR (see Figure \ref{fig:OPRconcept}). 

\subsection{D$_3^+$}\label{sec:D3+OPR}

D$_3^+$ has three spin configurations: ortho, para, and meta. In addition to OPR, we define ortho-to-meta ratio (hereafter, OMR) of D$_3^+$, $\text{OMR(D}_3^+{)}=n(\text{o-D}_3^+{)}/n(\text{m-D}_3^+{)}$. 
We assume that D$_3^+$ is formed through the reactions between D$_2$H$^+$ and HD, and destroyed through the reactions with H$_2$. The spin conversion proceeds via the reactions with HD and D$_2$ (see also Table \ref{table:OPRreaction_D3pOPR} and \ref{table:OPRreaction_D3pOMR}).  The time evolution equation for the D$_3^+$ OPR is given as

\begin{equation}\label{eq:OPRD3+}
     \begin{split}
         \frac{\mathrm{d}}{\mathrm{d} t} \text{OPR(D}_3^+{)}=&-\left[K_\mathrm{o\rightarrow p}+\frac{F(\text{p-D}_3^+)}{n_\text{op}(\text{D}_3^+)}+\frac{F_\mathrm{m\rightarrow p}}{n_\text{op}(\text{D}_3^+)}\right][\text{OPR(D}_3^+{)}]^2 \\
         &-\left[K_\mathrm{o\rightarrow p}-K_\mathrm{p\rightarrow o}-\frac{F(\text{o-D}_3^+)}{n_\text{op}(\text{D}_3^+)}\right. \\
         &\ \ \ \left.+\frac{F(\text{p-D}_3^+)}{n_\text{op}(\text{D}_3^+)}+ D(\text{o-D}_3^+)-D(\text{p-D}_3^+)\right. \\
         &\ \ \ \left.+\frac{F_\mathrm{m\rightarrow p}}{n_\text{op}(\text{D}_3^+)}-\frac{F_\mathrm{m\rightarrow o}}{n_\text{op}(\text{D}_3^+)}\right]\text{OPR(D}_3^+{)} \\
         &+K_\mathrm{p\rightarrow o}+\frac{F(\text{o-D}_3^+)}{n_\text{op}(\text{D}_3^+)}+\frac{F_\mathrm{m\rightarrow o}}{n_\text{op}(\text{D}_3^+)},
     \end{split}
\end{equation}
where 
\begin{equation}\label{eq:n_opD3+}
     \begin{split}
         n_\text{op}(\text{D}_3^+)=&n(\text{o-D}_3^+)+n(\text{p-D}_3^+) \\
         =&\frac{1+\text{OPR(D}_3^+{)}}{1+\text{OPR(D}_3^+{)}+\text{OPR(D}_3^+{)}/\text{OMR(D}_3^+{)}}n(\text{D}_3^+),
     \end{split}
\end{equation}
\begin{equation}\label{eq:Fmtop}
     \begin{split}
         F_\mathrm{m\rightarrow p}=&\left[k_\mathrm{m\rightarrow p}^\mathrm{oD_2}n(\text{o-D}_2)+k_\mathrm{m\rightarrow p}^\mathrm{pD_2}n(\text{p-D}_2)\right]n(\text{m-D}_3^+) \\
         =&\left[k_\mathrm{m\rightarrow p}^\mathrm{oD_2}\frac{\text{OPR}(\text{D}_2)}{1+\text{OPR}(\text{D}_2)}+k_\mathrm{m\rightarrow p}^\mathrm{pD_2}\frac{1}{1+\text{OPR}(\text{D}_2)}\right]n(\text{D}_2) \\
         &\frac{n(\text{D}_3^+)}{1+\text{OMR(D}_3^+{)}+\text{OMR(D}_3^+{)}/\text{OPR(D}_3^+{)}} \\
         =&\left[k_\mathrm{m\rightarrow p}^\mathrm{oD_2}f_\mathrm{o}(\text{D}_2)+k_\mathrm{m\rightarrow p}^\mathrm{pD_2}f_\mathrm{p}(\text{D}_2)\right]n(\text{D}_2) \\
         &\frac{n(\text{D}_3^+)}{1+\text{OMR(D}_3^+{)}+\text{OMR(D}_3^+{)}/\text{OPR(D}_3^+{)}},
     \end{split}
\end{equation}
and
\begin{equation}\label{eq:Fmtoo}
     \begin{split}
         F_\mathrm{m\rightarrow o}=&\left[k_\mathrm{m\rightarrow o}^\mathrm{oD_2}n(\text{o-D}_2)+k_\mathrm{m\rightarrow o}^\mathrm{pD_2}n(\text{p-D}_2)+k_\mathrm{m\rightarrow o}^\mathrm{HD}n(\text{HD})\right]n(\text{m-D}_3^+) \\
         =&\left[\left\{k_\mathrm{m\rightarrow o}^\mathrm{oD_2}\frac{\text{OPR}(\text{D}_2)}{1+\text{OPR}(\text{D}_2)}+k_\mathrm{m\rightarrow o}^\mathrm{pD_2}\frac{1}{1+\text{OPR}(\text{D}_2)}\right\}n(\text{D}_2)\right. \\
         &\left.+k_\mathrm{m\rightarrow o}^\mathrm{HD}n(\text{HD})\right]\frac{n(\text{D}_3^+)}{1+\text{OMR(D}_3^+{)}+\text{OMR(D}_3^+{)}/\text{OPR(D}_3^+{)}} \\
         =&\left[\left\{k_\mathrm{m\rightarrow o}^\mathrm{oD_2}f_\mathrm{o}(\text{D}_2)+k_\mathrm{m\rightarrow o}^\mathrm{pD_2}f_\mathrm{p}(\text{D}_2)\right\}n(\text{D}_2)+k_\mathrm{m\rightarrow o}^\mathrm{HD}n(\text{HD})\right] \\
         &\frac{n(\text{D}_3^+)}{1+\text{OMR(D}_3^+{)}+\text{OMR(D}_3^+{)}/\text{OPR(D}_3^+{)}}.
     \end{split}
\end{equation}
$k_\mathrm{m\rightarrow o}^\mathrm{HD}$ is the rate coefficient for the conversion of m-D$_3^+$ to o-D$_3^+$ via the reaction with HD:
\begin{equation}
    \text{m-D}_3^+ + \text{HD} \rightarrow \text{o-D}_3^+ + \text{HD},
\end{equation}
and $k_\mathrm{m\rightarrow o(p)}^\mathrm{o(p)D_2}$ is the rate coefficient for the conversion of m-D$_3^+$ to o(p)-D$_3^+$ via the reaction with o(p)-D$_2$:
\begin{equation}
    \text{m-D}_3^+ + \text{o(p)-D}_2 \rightarrow \text{o(p)-D}_3^+ + \text{o/p-H}_2.
\end{equation}
Similarly, the equation for the OMR D$_3^+$ is given as
\begin{equation}\label{eq:OMRD3+}
     \begin{split}
         \frac{\mathrm{d}}{\mathrm{d} t} \text{OMR(D}_3^+{)}=&-\left[K_\mathrm{o\rightarrow m}+\frac{F(\text{m-D}_3^+)}{n_\text{om}(\text{D}_3^+)}+\frac{F_\mathrm{p\rightarrow m}}{n_\text{om}(\text{D}_3^+)}\right][\text{OMR(D}_3^+{)}]^2 \\
         &-\left[K_\mathrm{o\rightarrow m}-K_\mathrm{m\rightarrow o}-\frac{F(\text{o-D}_3^+)}{n_\text{om}(\text{D}_3^+)}\right. \\
         &\ \ \ \left.+\frac{F(\text{m-D}_3^+)}{n_\text{om}(\text{D}_3^+)}+ D(\text{o-D}_3^+)-D(\text{m-D}_3^+)\right. \\
         &\ \ \ \left.+\frac{F_\mathrm{p\rightarrow m}}{n_\text{om}(\text{D}_3^+)}-\frac{F_\mathrm{p\rightarrow o}}{n_\text{om}(\text{D}_3^+)}\right]\text{OMR(D}_3^+{)} \\
         &+K_\mathrm{m\rightarrow o}+\frac{F(\text{o-D}_3^+)}{n_\text{om}(\text{D}_3^+)}+\frac{F_\mathrm{p\rightarrow o}}{n_\text{om}(\text{D}_3^+)},
     \end{split}
\end{equation}
where

\begin{equation}\label{eq:n_omD3+}
     \begin{split}
         n_\text{om}(\text{D}_3^+)=&n(\text{o-D}_3^+)+n(\text{m-D}_3^+) \\
         =&\frac{1+\text{OMR(D}_3^+{)}}{1+\text{OMR(D}_3^+{)}+\text{OMR(D}_3^+{)}/\text{OPR(D}_3^+{)}}n(\text{D}_3^+),
     \end{split}
\end{equation}
\begin{equation}\label{eq:Fptom}
     \begin{split}
         F_\mathrm{p\rightarrow m}=&\left[k_\mathrm{p\rightarrow m}^\mathrm{oD_2}n(\text{o-D}_2)+k_\mathrm{p\rightarrow m}^\mathrm{pD_2}n(\text{p-D}_2)\right]n(\text{p-D}_3^+) \\
         =&\left[k_\mathrm{p\rightarrow m}^\mathrm{oD_2}f_\mathrm{o}(\text{D}_2)+k_\mathrm{p\rightarrow m}^\mathrm{pD_2}f_\mathrm{p}(\text{D}_2)\right]n(\text{D}_2) \\
         &\frac{n(\text{D}_3^+)}{1+\text{OPR(D}_3^+{)}+\text{OPR(D}_3^+{)}/\text{OMR(D}_3^+{)}},
     \end{split}
\end{equation}
and
\begin{equation}\label{eq:FptooD3+}
     \begin{split}
         F_\mathrm{p\rightarrow o}=&\left[k_\mathrm{p\rightarrow o}^\mathrm{oD_2}n(\text{o-D}_2)+k_\mathrm{p\rightarrow o}^\mathrm{pD_2}n(\text{p-D}_2)+k_\mathrm{p\rightarrow o}^\mathrm{HD}n(\text{HD})\right]n(\text{p-D}_3^+) \\
         =&\left[\left\{k_\mathrm{p\rightarrow o}^\mathrm{oD_2}f_\mathrm{o}(\text{D}_2)+k_\mathrm{p\rightarrow o}^\mathrm{pD_2}f_\mathrm{p}(\text{D}_2)\right\}n(\text{D}_2)+k_\mathrm{p\rightarrow o}^\mathrm{HD}n(\text{HD})\right] \\
         &\frac{n(\text{D}_3^+)}{1+\text{OPR(D}_3^+{)}+\text{OPR(D}_3^+{)}/\text{OMR(D}_3^+{)}}.
     \end{split}
\end{equation}
$k_\mathrm{p\rightarrow o}^\mathrm{HD}$ is the rate coefficient for the conversion of p-D$_3^+$ to o-D$_3^+$ via the reaction with HD:
\begin{equation}
    \text{p-D}_3^+ + \text{HD} \rightarrow \text{o-D}_3^+ + \text{HD},
\end{equation}
and $k_\mathrm{p\rightarrow o(m)}^\mathrm{o(p)D_2}$ is the rate coefficient for the conversion of p-D$_3^+$ to o(m)-D$_3^+$ via the reaction with o(p)-D$_2$:
\begin{equation}
    \text{p-D}_3^+ + \text{o(p)-D}_2 \rightarrow \text{o(m)-D}_3^+ + \text{o/p-H}_2.
\end{equation}

We solve equations (\ref{eq:OPRD3+}) and (\ref{eq:OMRD3+}) with the quasi-steady-state assumption, i.e., $\mathrm{d}\text{OPR(D}_3^+{)}/\mathrm{d}t=\mathrm{d}\text{OMR(D}_3^+{)}/\mathrm{d}t=0$. Since $\text{OMR(D}_3^+{)}$ appears in equation (\ref{eq:OPRD3+}) and $\text{OPR(D}_3^+{)}$ appears in equation (\ref{eq:OMRD3+}), we solve the two equations iteratively at a given time.   

\subsection{D$_2$}\label{sec:D2OPR}

As in the case of H$_2$, we calculate the total formation and destruction rate of D$_2$ via gas-phase and grain surface reactions to evaluate the D$_2$ OPR. When D$_2$ is formed, the D$_2$ OPR is assumed to be the statistical value of two, as for the H$_2$ formation. However, similar to the case of the H$_2$ formation rates, several reactions constitute exceptions. The proton transfer reactions between o-D$_2$H$^+$ and neutral species (e.g., CO, O, N$_2$, OH, and H$_2$O) and the dissociative recombination reactions between o-D$_2$H$^+$ and electrons or dust grains form o-D$_2$, and the reactions between p-D$_2$H$^+$ and those chemical species form p-D$_2$. Similarly, the rates of the reactions between D$_3^+$ and the species listed above to form D$_2$ also depend on the spin states of D$_3^+$. We calculate the total formation rates of o/p-D$_2$ by weighing the reaction rates using the D$_2$H$^+$ and D$_3^+$ OPRs and D$_3^+$ OMR. In addition to the formation or destruction reactions of D$_2$ listed in Table \ref{table:OPRreaction_D2}, we also include reactions between isotopologues of H$_2$ and H$_3^+$ that contribute to the formation or destruction of D$_2$ \citep{Hily-Blant_2018}. We calculate the D$_2$ OPR with the quasi-steady-state assumption.

\section{Test calculations}\label{sec:testcalc}

\begin{figure*}
    \centering
    \includegraphics[width=\hsize]{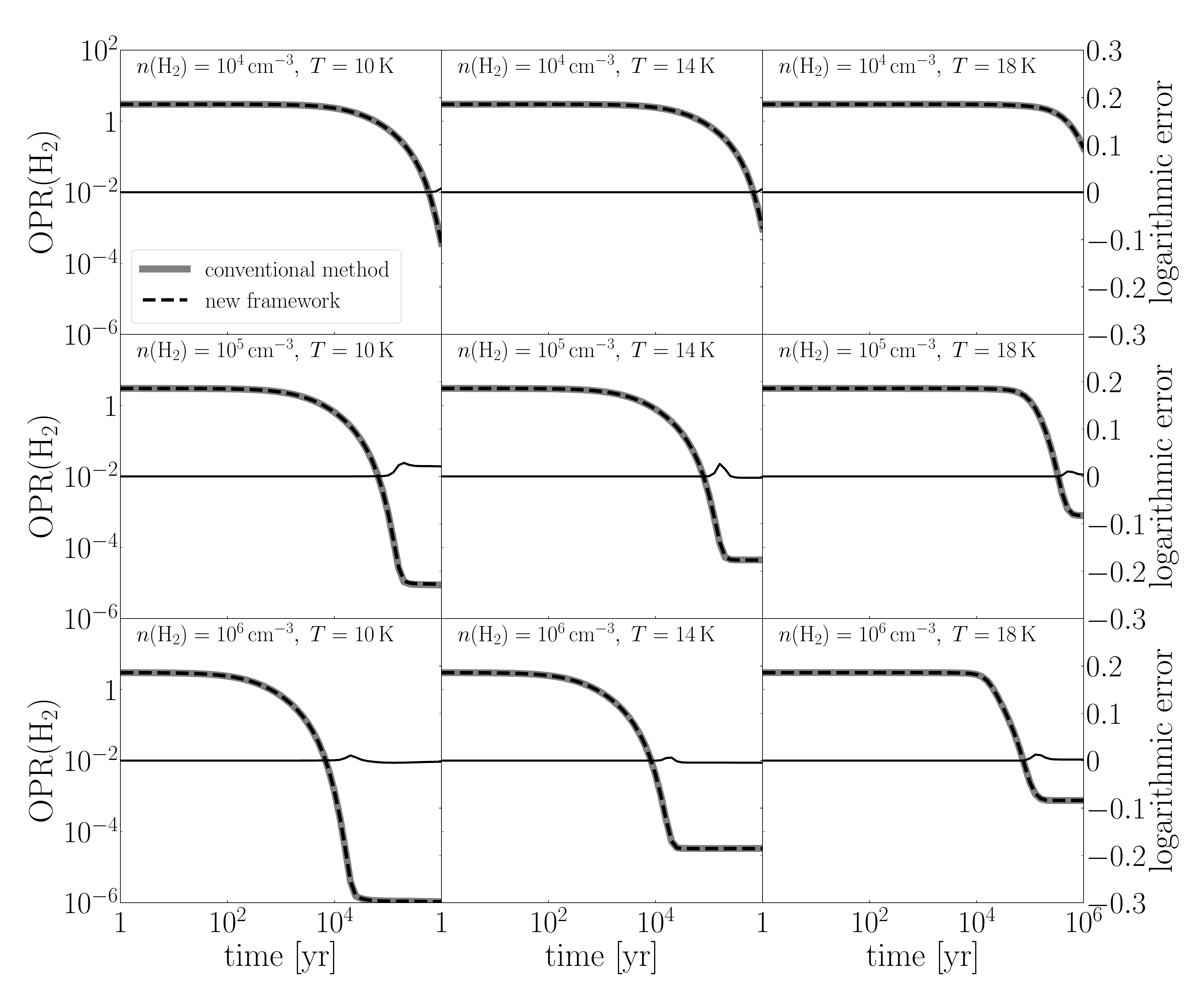}
    \caption{Time evolution of the H$_2$ OPR calculated by the conventional method (solid gray lines) and the new framework (dashed black lines) in the static model under various physical conditions. The thinner solid black lines show the logarithmic error between the new framework and the conventional method.}
    \label{fig:OPRH2}
\end{figure*}

\begin{figure*}
    \centering
    \includegraphics[width=\hsize]{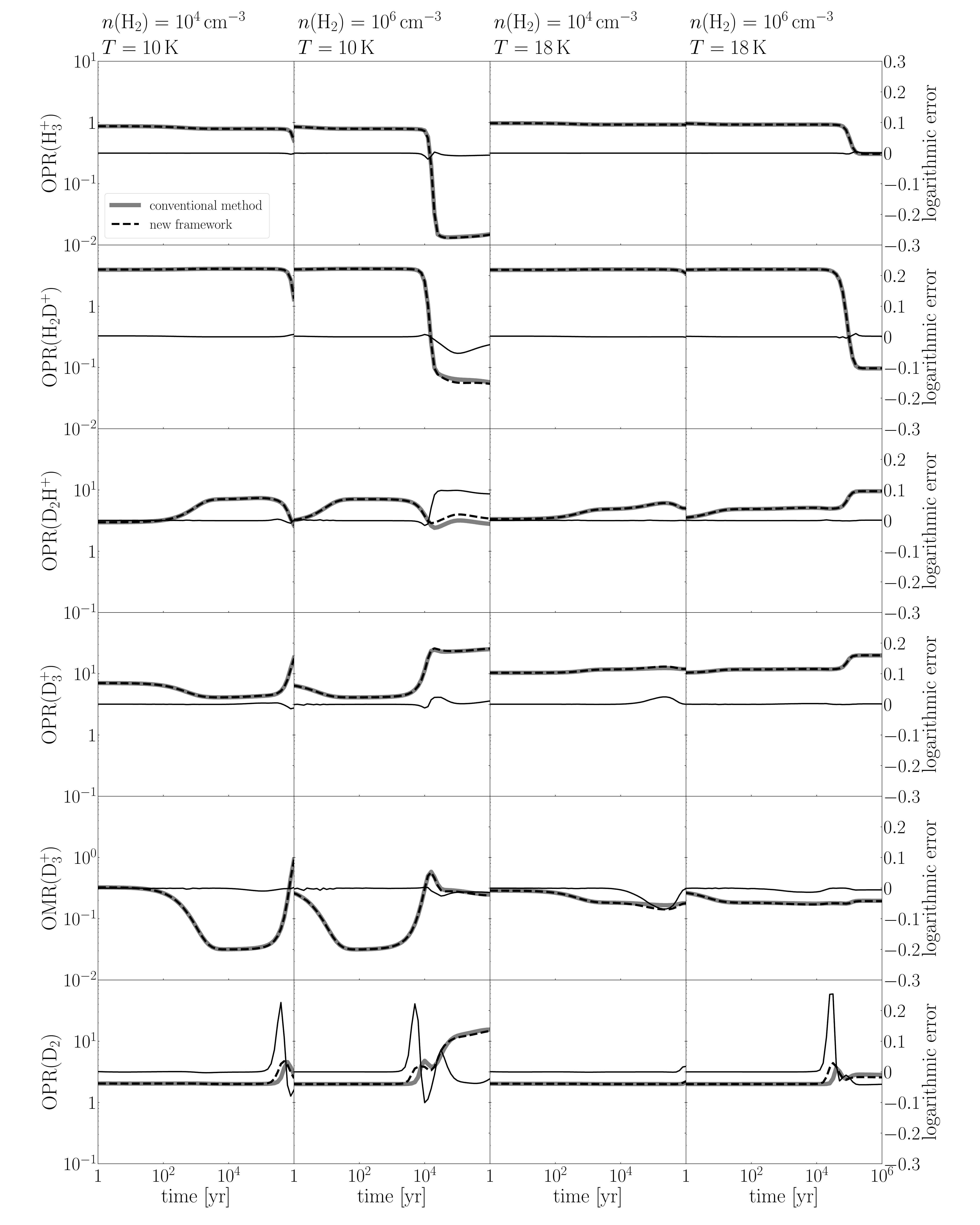}
    \caption{The OPR or OMR of H$_3^+$ isotopologues and D$_2$ calculated by the conventional method (solid gray lines) and the new framework (dashed black lines) in the static model. The thinner solid black lines show the logarithmic error between the new framework and the conventional method.}
    \label{fig:OPRH3+}
\end{figure*}

\begin{figure}[h!]
    \centering
    \includegraphics[width=0.9\hsize]{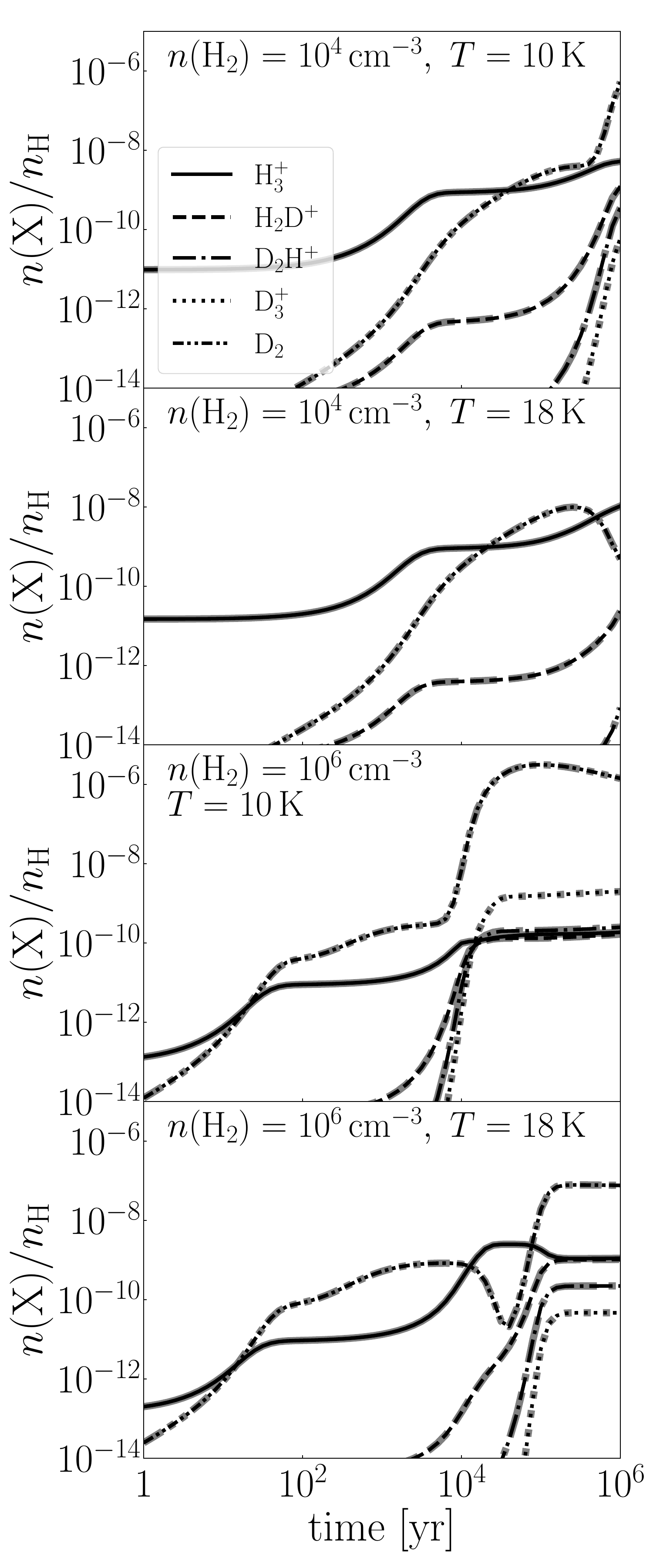}
    \caption{Time evolution of the fractional abundances of H$_3^+$, its isotopologues, and D$_2$ calculated by the conventional method (thick gray lines) and the new framework (thin black lines) in the static model under various physical conditions.}
    \label{fig:H3pab}
\end{figure}

\begin{figure}[h!]
    \centering
    \includegraphics[width=0.9\hsize]{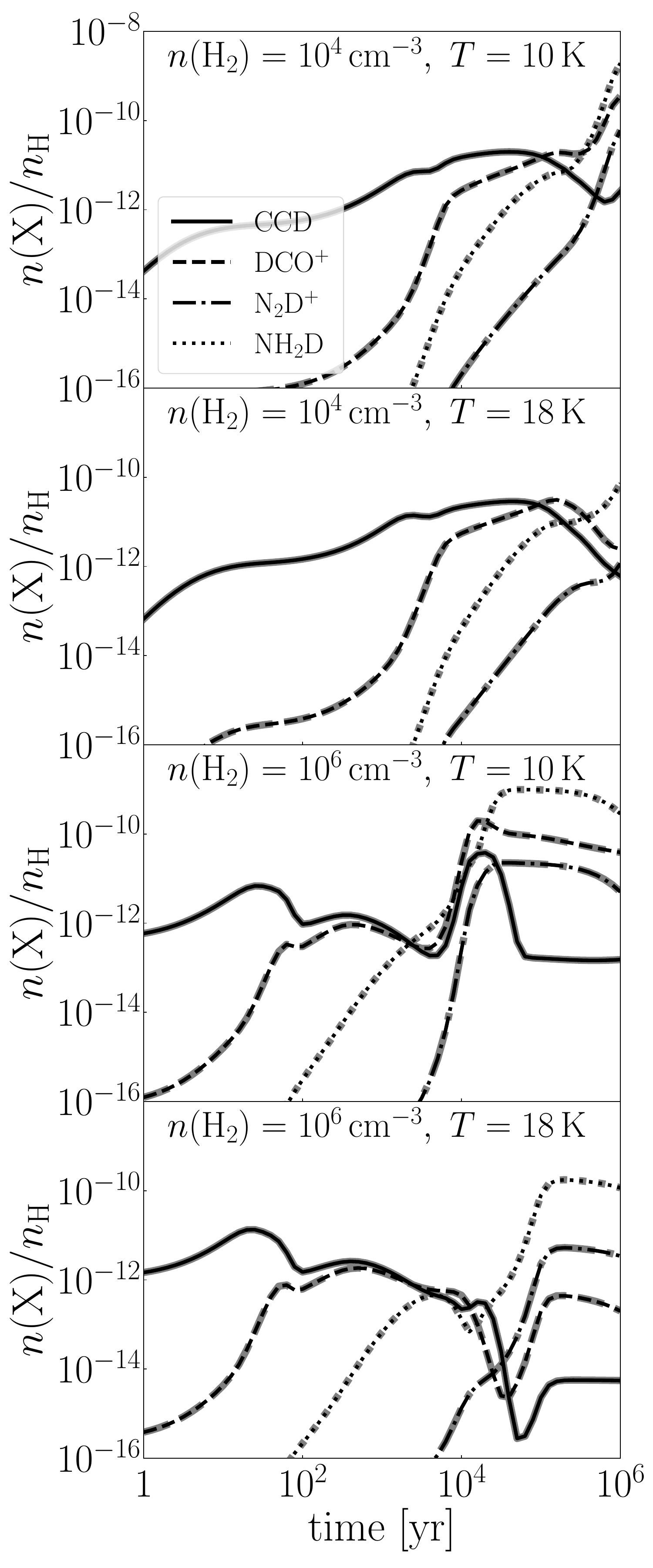}
    \caption{Similar to Figure \ref{fig:H3pab}, but for CCD, DCO$^+$, N$_2$D$^+$, and NH$_2$D.}
    \label{fig:Dspab}
\end{figure}

\begin{figure}[h!]
    \centering
    \includegraphics[width=\hsize]{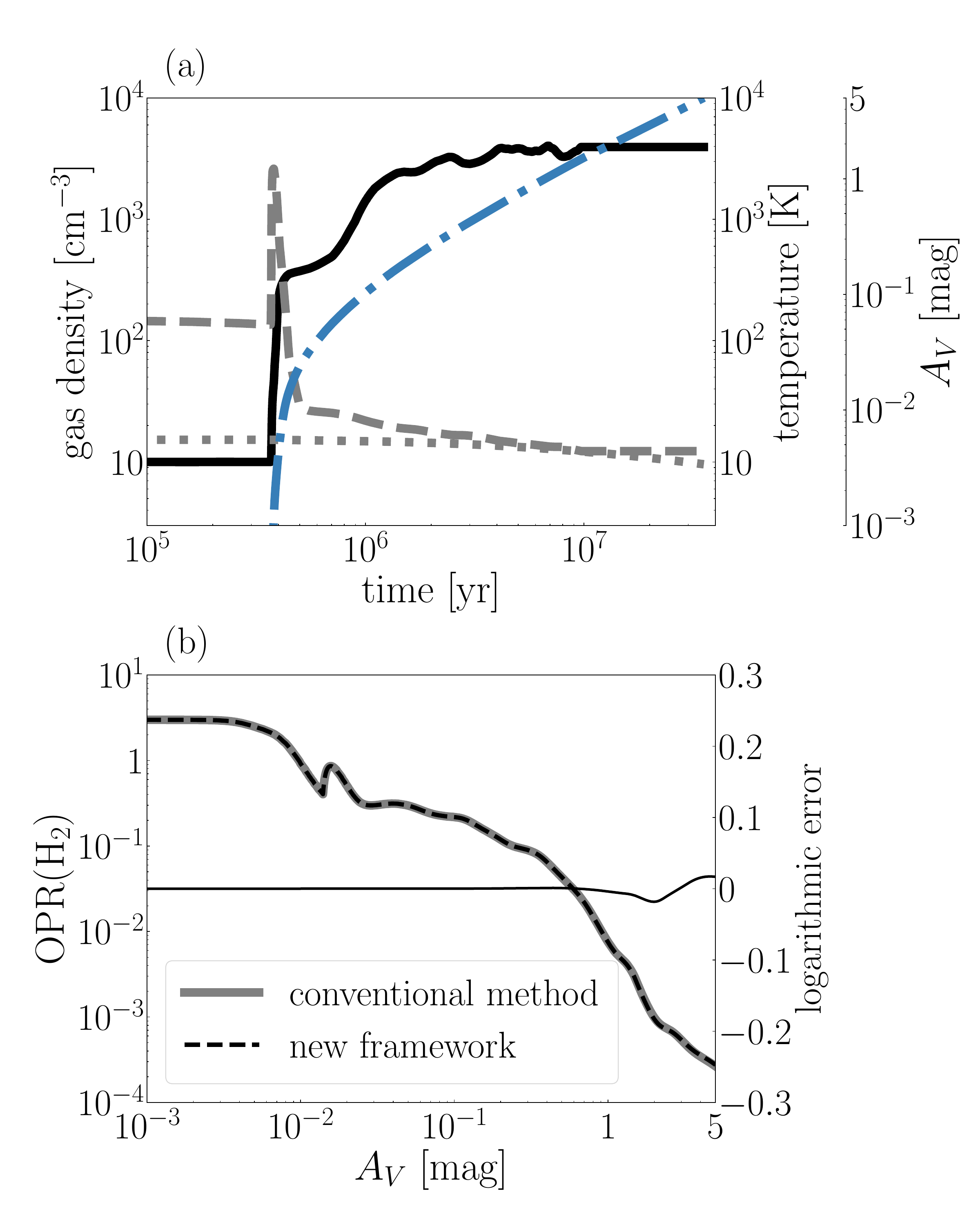}
    \caption{(a) The time evolution of gas density (solid black line), gas temperature (dashed gray line), dust temperature (dotted gray line), and visual extinction (dash-dotted blue line) of the dynamical model. (b) H$_2$ OPR as a function of $A_V$ calculated by the conventional method (solid gray lines) and the new framework (dashed black lines) in the dynamical model. The thinner solid black line shows the logarithmic error between the new framework and the conventional method.}
    \label{fig:OPRH2_1d}
\end{figure}

\begin{figure}[h!]
    \centering
    \includegraphics[width=\hsize]{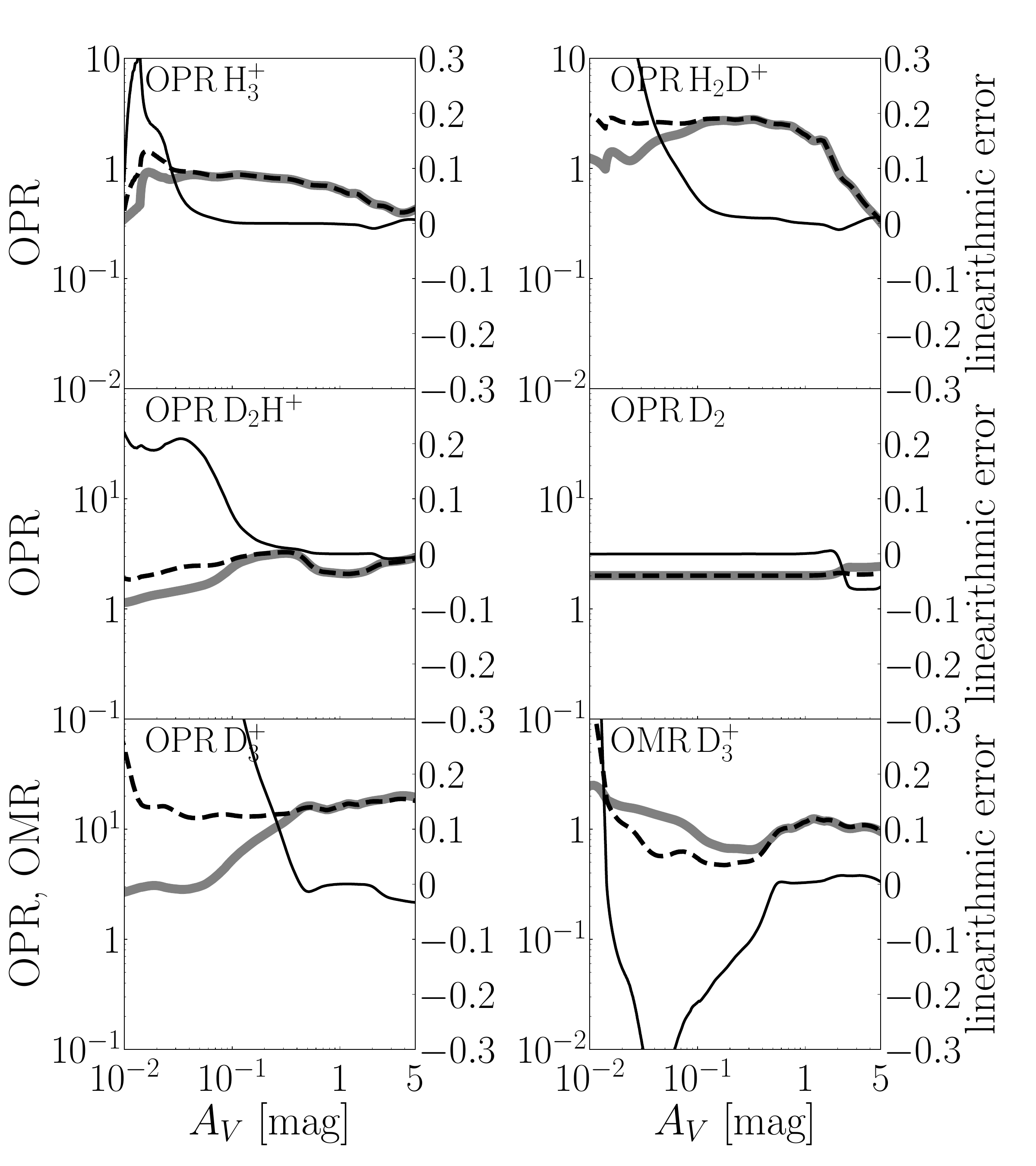}
    \caption{The OPR or OMR of H$_3^+$ isotopologues and D$_2$ as functions of $A_V$ calculated by the conventional method (solid gray lines) and the new framework (dashed black lines) in the dynamical model. The thinner solid black lines show the logarithmic error between the new framework and the conventional method.}
    \label{fig:OPRH3+_1d}
\end{figure}

\begin{figure}[h!]
    \centering
    \includegraphics[width=\hsize]{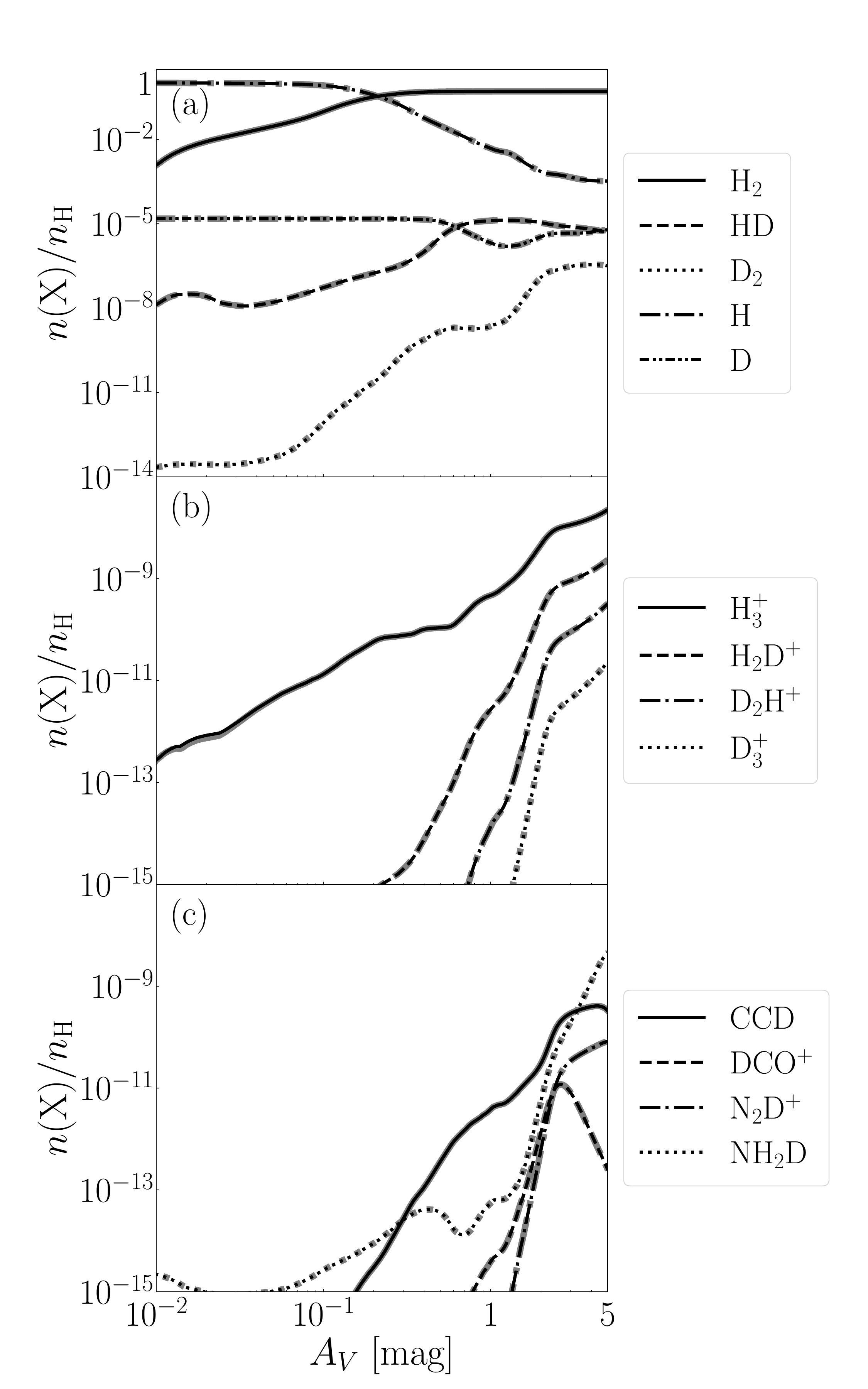}
    \caption{fractional abundances of (a) H$_2$, H, and their isotopes, (b) H$_3^+$ and its isotopes, and (c) representative deuterated molecules as a function of $A_V$ calculated by the conventional method (thick gray lines) and the new framework (thin black lines) in the dynamical model.}
    \label{fig:OPRH2_1d_frac}
\end{figure}

\subsection{Calculation setup}\label{sec:setting}
We implement the framework described in the previous section into {\sc rokko}, a chemical network calculation code presented in \citet{Furuya_2015}. The calculation setup, except for the physical and initial conditions, follows \citet{Furuya_2026}. We adopted the three-phase model that considers the gas phase, grain surface, and inert ice mantle following \citet{Hasegawa_1993b}. The rate coefficients for reactions between $\text{H}_2$ and $\text{H}_3^+$ are taken from \citet{Hily-Blant_2018} (see also \citet{Sipila_2017}). The rate coefficient of the reaction between o- or p-H$_2$ and proton is adopted from \citet{Honvault_2011,Honvault_2012}. We incorporate the nuclear spin conversion rate of H$_2$ on grain surfaces following \citet{Furuya_2019,Furuya_2026}, assuming the energy difference between o- and p-H$_2$ to be 170.5\,K. See \citet{Furuya_2015} and \citet{Furuya_2026} for further details. To test the accuracy of our framework, we conducted two types of calculations: in the first calculation, following the conventional method, the individual spin states of H$_2$, H$_3^+$, and their isotopologues are treated separately in the chemical network as in \citet{Furuya_2026}. In the second calculation, following our new framework, the OPRs and OMR are calculated using the OPR and OMR equations, and the total fractional abundances of all spin states are calculated with the chemical network in which reactions are not separated by the spin states. We consider the chemical reactions listed in Table \ref{table:OPRreaction} to \ref{table:OPRreaction_D2} to solve the OPR and OMR equations.

We adopted two sets of physical conditions for the test calculation. One is the static model with dark cloud conditions, in which the physical quantities (gas density $n_\mathrm{H}$, gas and dust temperature $T$, and visual extinction $A_V$) are assumed to be constant with time. We vary the gas density from $2\times 10^4\,\mathrm{cm^{-3}}$ to $2\times 10^6\,\mathrm{cm^{-3}}$, and the temperature from $10\,\mathrm{K}$ to $18\,\mathrm{K}$. The gas temperature is assumed to be the same as the dust temperature. The visual extinction is set to $10\,\mathrm{mag}$. The other is the dynamical model with molecular cloud formation conditions, where the physical quantities vary with time. We use the fiducial model (N10V10sin020) of \citet{Komichi_2024}, which is a one-dimensional mean-flow model derived from the three-dimensional magnetohydrodynamics simulation of converging flow of atomic gas. Panel (a) of Figure \ref{fig:OPRH2_1d} shows the temporal variation of gas density, gas and dust temperature, and visual extinction of the dynamical model.

The elemental abundances are taken from \citet{Aikawa_1999}. Hydrogen and deuterium are initially assumed to exist as H$_2$ and HD, respectively, in the static model, whereas they are initially atomic in the dynamical model. The other elements are initially neutral atoms or ions in both models, depending on their ionization potential. The initial H$_2$ OPR is set to three in the static models. We adopt the strength of the interstellar radiation field of Draine field \citep{Draine_1978}, and the cosmic-ray ionization rate of $1.3\times 10^{-17}\,\mathrm{s^{-1}}$.

\subsection{Results}\label{sec:result}

Figure \ref{fig:OPRH2} shows the time evolution of the H$_2$ OPR in the static model. The timescale of the spin state conversion depends on both gas density and temperature. The results calculated by the conventional method and our new framework are nearly identical. In order to evaluate the accuracy of our framework quantitatively, we calculated the logarithmic errors, $\Delta(\text{OPR(H}_2\text{)})=\log_{10}\,[\text{OPR(H}_2\text{)}^\text{new}/\text{OPR(H}_2\text{)}^\text{con}]$, between the two methods at each timestep. The logarithmic error is up to 0.03, which means that the error of our framework is within 7\%. This error is acceptable since it is within the typical error of the chemical network itself \citep{Sipila_2017}. We note that this result is natural because the OPR equation for H$_2$ originates from the rate equations for ortho and para-H$_2$, which are solved in the conventional method. The only approximations introduced in our framework are the simplification of the spin conversion rate (equation (\ref{eq:OPRH2_conversionrate})) and that H$_2$ is mostly present in the gas phase rather than on grains (equation (\ref{eq:totoH2_})).

Figure \ref{fig:OPRH3+} shows the calculated OPRs or OMR of H$_3^+$ isotopologues and the OPR of D$_2$. The solid black lines in Figure \ref{fig:OPRH3+} demonstrate the logarithmic errors of the new framework. The calculated OPR and OMR of H$_3^+$ and its isotopologues with our framework show lower than 0.1 of the logarithmic error, which demonstrates reasonable agreement with those obtained by the conventional method.  OPR of D$_2$ with our framework evolves slightly earlier than that obtained with the conventional method. We suspect this is due to the non-equilibrium effect. Since the D$_2$ OPR is mainly determined by the total formation rate of o- and p-D$_2$, and the formation timescale of D$_2$ can be long enough, the D$_2$ OPR takes time to reach its quasi-steady-state value at each moment, while our framework assumes that the D$_2$ OPR immediately reaches the quasi-steady-state value. However, the error of D$_2$ OPR is still within about a factor of two. Overall, our framework also gives accurate results for OPR or OMR of H$_3^+$, its isotopologues, and D$_2$ in the static case.

Figure \ref{fig:H3pab} and \ref{fig:Dspab} show the fractional abundances of H$_3^+$, its isotopologues, and deuterated molecules calculated with the conventional method and the new framework. Although there is a possible error of up to twice in the OPR and OMR in our framework (Figure \ref{fig:OPRH3+}), the fractional abundances of the molecules are consistent between the two methods. This demonstrates that the accuracy of OPR and OMR in our framework is sufficient to solve the fractional abundances of molecules that have spin isomers and deuterated molecules.

Panel (b) of Figure \ref{fig:OPRH2_1d} displays the H$_2$ OPR in the dynamical model calculated with our framework and the conventional method. Similar to the static model, our framework gives consistent results for the H$_2$ OPR, even when physical quantities vary as a function of time. 

Figure \ref{fig:OPRH3+_1d} shows the OPR or OMR of H$_3^+$ and its isotopologues, as well as the OPR of D$_2$. The OPRs of H$_3^+$ and H$_2$D$^+$ show reasonable agreement at $A_V\gtrsim 0.1\,\mathrm{mag}$, while the OPR and OMR of H$_3^+$ isotopologues agree at $A_V\gtrsim 1\,\mathrm{mag}$. On the other hand, deviations appear at lower $A_V$. This is because of the lower H$_2$ abundance than H atoms, and the lower HD abundance than D atoms (panel (a) of Figure \ref{fig:OPRH2_1d_frac}). When H$_2$ and HD abundances are low, minor reaction channels that are not included in our framework mainly contribute to their formation, which results in larger uncertainty of H$_3^+$, H$_2$D$^+$, and D$_2$H$^+$ OPRs at $A_V\lesssim 0.1\,\mathrm{mag}$ and of D$_3^+$ OPR and OMR at $A_V\lesssim 0.5\,\mathrm{mag}$. 
The calculated D$_2$ OPR shows a slight difference from the results of the chemical network. This is because the proton transfer reactions between D$_2$H$^+$ and neutral molecules such as HCN and CN, which are not included as exceptional reactions in the calculation of the D$_2$ formation rate (see Section \ref{sec:D2OPR}), become important. However, the logarithmic error is still within 0.1, and our framework still gives accurate results.
We note that, similar to the case of the static model, the total abundances of all spin states of each molecule and the abundances of deuterated molecules calculated with the two methods are almost identical (Figure \ref{fig:OPRH2_1d_frac}). 

\section{Discussion}\label{sec:discussion}

\subsection{Limitation}\label{sec:limit_of_framework}

As for H$_2$, our framework can reproduce the OPR consistent with the conventional method under the physical conditions adopted in this work. In the higher-density conditions ($\gtrsim 10^6\,\mathrm{cm^{-3}}$, e.g., protoplanetary disks), where H$_2$D$^+$ and D$_2$H$^+$ could be more abundant, in addition to H$_3^+$, they would also contribute to the exchange of the spin states of H$_2$, which is not included in this work. However, in such dense conditions ($\gtrsim 10^4\,\mathrm{cm^{-3}}$), the spin conversion rate is dominated by the NSC on grain surfaces \citep{Furuya_2026}. Given that H$_2$D$^+$ and D$_2$H$^+$ are less abundant than H$_3^+$ \citep{Walmsley_2004}, the uncertainty of our framework is expected to remain small. We also note that our framework is still accurate enough in the case of initially lower H$_2$ OPR or the case of higher gas and dust temperature ($T=30\,\mathrm{K}$).

As for H$_3^+$, its isotopologues, and D$_2$, the uncertainty associated with our framework may become larger in the case of dense conditions. In this work, we selected some relevant chemical reactions to solve OPR and OMR of H$_3^+$, its isotopologues, and D$_2$. We basically chose these reactions listed in Tables \ref{table:OPRreaction} to \ref{table:OPRreaction_D2} as they contribute about more than 10\% of the total formation or destruction rates of each molecule. We omit some reactions contributing more than 10\% of the formation or destruction rates (e.g., H$_3^+$ + D$_2$ reaction system for the destruction of H$_3^+$), because the calculations sometimes become heavy and fragile when we include the reactions.  As we demonstrate in Section \ref{sec:result}, our current framework is accurate enough for solving the time evolution of the fractional abundances of deuterated molecules with the selected physical conditions. However, such reactions can be important in high-density conditions, where D$_2$ tends to be abundant. As for D$_2$, some neutral molecules which are not considered in our current framework (see Section \ref{sec:D2OPR}) might contribute to D$_2$ formation. However, as long as the quasi-steady-state assumption is valid, our framework would give accurate results by additionally considering such reactions even in the case of dense conditions.

The quasi-steady-state assumption is valid when the timescale of ion-molecule reactions between H$_3^+$ or its isotopologues and H$_2$ or HD is shorter than the conversion timescale of H$_2$ OPR. The typical timescales of ion-molecule reactions for H$_3^+$ and its isotopologues, $\tau_\mathrm{im}$, is
\begin{equation}
    \tau_\mathrm{im} \simeq 3\times 10^{-2}\,\mathrm{yr} \left(\frac{n(\mathrm{H_2})}{10^4\,\mathrm{cm^{-3}}}\right)^{-1}
\end{equation}
with H$_2$, and
\begin{equation}
    \tau_\mathrm{im} \simeq 3\times 10^{3}\,\mathrm{yr} \left(\frac{n_\mathrm{H}}{10^4\,\mathrm{cm^{-3}}}\right)^{-1}\left(\frac{a(\mathrm{HD})}{10^{-5}}\right)^{-1}
\end{equation}
with HD, where $n(\mathrm{H_2})$ is the H$_2$ number density, $n_\mathrm{H}$ is the total hydrogen nuclei number density, and $a(\mathrm{X})$ is the fractional abundance of X with respect to the total hydrogen nuclei. On the other hand, the timescale of the variation of H$_2$ OPR, $\tau_\mathrm{OPR}$, is roughly given as
\begin{equation}
    \tau_\mathrm{OPR} \gtrsim 5\times 10^5\,\mathrm{yr} \left(\frac{n_\mathrm{H}}{10^4\,\mathrm{cm^{-3}}}\right)^{-1}
\end{equation}
when OPR changes mainly due to nuclear spin conversions on grain surfaces \citep{Furuya_2019,Furuya_2026}, and
\begin{equation}
    \tau_\mathrm{OPR} \gtrsim 10^6\,\mathrm{yr} \left(\frac{n_\mathrm{H}}{10^4\,\mathrm{cm^{-3}}}\right)^{-1}\left(\frac{a(\mathrm{H^+})}{10^{-8}}\right)^{-1}
\end{equation}
when OPR changes mainly due to the gas-phase reaction with H$^+$. Therefore, the steady-state approximation should be valid in dense conditions. In addition, if the physical quantities vary with time, the physical timescale should be longer than the timescales of the ion-molecule reactions. For example, the free-fall timescale, $\tau_\mathrm{ff}$ expressed as
\begin{equation}
    \tau_\mathrm{ff}\simeq 3\times 10^5\mathrm{yr}\,\left(\frac{n_\mathrm{H}}{10^4\,\mathrm{cm^{-3}}}\right)^{-1/2}
\end{equation}
can be about ten times longer than $\tau_\mathrm{im}$ at $n_\mathrm{H}>10^2\,\mathrm{cm^{-3}}$.

\subsection{Advantages of our framework}\label{sec:benefit}

One of the advantages of our framework is that it can simplify the chemical network. Basically, almost all reactions do not significantly affect, or are not significantly affected by, the OPRs or OMR. However, with the conventional method, we have to treat the reactions involving H$_2$, H$_3^+$, and their isotopologues separately with their spin states. Such a treatment increases the total number of reactions in the chemical network. Recently, the chemical network become increasingly complex. For example, \citet{Ichimura_2026} includes a carbon isotope, ${}^{13}$C, and the position of the ${}^{13}$C atom within each molecule in their chemical network, which results in about $6\times 10^4$ reactions in total. Including the deuterium chemistry further increases the total number of reactions by more than a factor of five, which would make such chemical networks difficult to handle. Our framework can reduce the number of reactions and contribute to the simplification of the chemical network. For the setup described in Section \ref{sec:setting}, our framework reduces the total number of reactions from about 72,000 in the conventional method to about 54,000. By adding OPR equations in numerical codes for solving the chemical network, we can take into account the OPR without modifying the chemical network itself. For example, the new framework would make it easier for OPRs to be solved self-consistently in the hydrodynamics simulations that include a chemical network \citep[e.g.,][]{Bovion_2019,Bovino_2021,Lupi_2021}. 

On the other hand, our framework does not necessarily shorten the calculation time compared to the conventional method. The computational bottleneck in the chemical network calculations in general is the evaluation of the Jacobian matrix of the simultaneous ordinary differential equations. The computational cost basically depends on the total number of chemical species. Generally, the calculation time can be shorter when we decrease the number of species and reactions. However, when we use our framework only for ortho and para-H$_2$, we include the H$_2$ OPR as an additional variable instead of including ortho and para-H$_2$ separately, and the total number of species does not change significantly. If we solve OPR or OMR of D$_2$, H$_3^+$, and its isotopologues with the steady-state assumption, we can reduce the number of species by six from the original chemical network. However, when more than 10$^2$ species are included in the network, the effect of the reduction on computational cost would be negligible.

We measure the calculation time during our test calculations. When we adopt the static model, in the case of $n(\mathrm{H})=2\times 10^5\,\mathrm{cm^{-3}}$ and $T=10\,\mathrm{K}$, the calculation time with our framework is about 79.0\% of that of the conventional method. In the case of $n(\mathrm{H})=2\times 10^6\,\mathrm{cm^{-3}}$ and $T=10\,\mathrm{K}$, the calculation time with our framework is about 51.1\% of that of the original network. When we adopt the dynamical model, the calculation time with our framework was about two times longer than with the conventional method. The calculation time is also determined by how the solutions converge during the iteration, which depends on the physical conditions. Our framework at least does not significantly increase or decrease the calculation time. 

\section{Conclusion}\label{sec:summary}

In this paper, we developed a new framework to calculate the OPRs or OMR of H$_2$ and H$_3^+$ isotopologues in astrochemical models. Unlike the conventional method, which explicitly separates chemical reactions according to nuclear spin configurations, our framework solves the time evolution equations for the H$_2$ OPR and the total fractional abundances of all spin states of each molecule. As for H$_3^+$ isotopologues and D$_2$, we adopt the quasi-steady-state approximation to solve their OPR and OMR equations at each timestep. This approximation is valid if the timescale of the variation of their OPR and OMR is shorter than that of the H$_2$ OPR and physical quantities. We conducted test calculations using our framework in both static and dynamical physical conditions. The results of the test calculations, including the total abundances summed over all nuclear spin states, show that the typical logarithmic error of our framework against the conventional methods is up to 0.1, which validates the accuracy of our framework. Our framework can include nuclear spin chemistry of H$_2$ and H$_3^+$ isotopologues in any chemical network models in which reactions are not separated according to nuclear spin isomers, which makes it simple to handle the nuclear spin chemistry in astrochemical models. Besides the conventional method, our framework is an alternative approach for investigating the nuclear spin chemistry in star-forming regions.

\begin{acknowledgements}
      Y. K. thanks Yuri Aikawa and Kengo Tomida for fruitful discussions. The calculations were carried out on the general-purpose PC cluster at the Center for Computational Astrophysics, National Astronomical Observatory of Japan. This work is supported by JSPS KAKENHI Grant Numbers JP24KJ0901 and JP25K07364. Y. K. acknowledges the support by International Graduate Program for Excellence in Earth-Space Science (IGPEES) of the University of Tokyo.
\end{acknowledgements}

%

\bibliographystyle{aa}
\bibliography{citation}

@ARTICLE{Boley_2007,
       author = {{Boley}, Aaron C. and {Hartquist}, Thomas W. and {Durisen}, Richard H. and {Michael}, Scott},
        title = "{The Internal Energy for Molecular Hydrogen in Gravitationally Unstable Protoplanetary Disks}",
      journal = {\apjl},
         year = 2007,
        month = feb,
       volume = {656},
       number = {2},
        pages = {L89-L92},
          doi = {10.1086/512235},
archivePrefix = {arXiv},
       eprint = {astro-ph/0701146},
 primaryClass = {astro-ph},
       adsurl = {https://ui.adsabs.harvard.edu/abs/2007ApJ...656L..89B}
}

@ARTICLE{Decampli_1978,
       author = {{Decampli}, W.~M. and {Cameron}, A.~G.~W. and {Bodenheimer}, P. and {Black}, D.~C.},
        title = "{Ortho- and para-hydrogen in dense clouds, protoplanets, and planetary atmospheres.}",
      journal = {\apj},
         year = 1978,
        month = aug,
       volume = {223},
        pages = {854-858},
          doi = {10.1086/156318},
       adsurl = {https://ui.adsabs.harvard.edu/abs/1978ApJ...223..854D}
}

@ARTICLE{Vaytet_2014,
       author = {{Vaytet}, Neil and {Tomida}, Kengo and {Chabrier}, Gilles},
        title = "{On the role of the H$_{2}$ ortho:para ratio in gravitational collapse during star formation}",
      journal = {\aap},
         year = 2014,
        month = mar,
       volume = {563},
          eid = {A85},
        pages = {A85},
          doi = {10.1051/0004-6361/201322855},
archivePrefix = {arXiv},
       eprint = {1401.4299},
 primaryClass = {astro-ph.SR},
       adsurl = {https://ui.adsabs.harvard.edu/abs/2014A&A...563A..85V}
}

@ARTICLE{Glover_2008,
       author = {{Glover}, S.~C.~O. and {Abel}, T.},
        title = "{Uncertainties in H$_{2}$ and HD chemistry and cooling and their role in early structure formation}",
      journal = {\mnras},
         year = 2008,
        month = aug,
       volume = {388},
       number = {4},
        pages = {1627-1651},
          doi = {10.1111/j.1365-2966.2008.13224.x},
archivePrefix = {arXiv},
       eprint = {0803.1768},
 primaryClass = {astro-ph},
       adsurl = {https://ui.adsabs.harvard.edu/abs/2008MNRAS.388.1627G}
}

@ARTICLE{Furuya_2015,
       author = {{Furuya}, K. and {Aikawa}, Y. and {Hincelin}, U. and {Hassel}, G.~E. and {Bergin}, E.~A. and {Vasyunin}, A.~I. and {Herbst}, E.},
        title = "{Water deuteration and ortho-to-para nuclear spin ratio of H$_{2}$ in molecular clouds formed via the accumulation of H I gas}",
      journal = {\aap},
         year = 2015,
        month = dec,
       volume = {584},
          eid = {A124},
        pages = {A124},
          doi = {10.1051/0004-6361/201527050},
archivePrefix = {arXiv},
       eprint = {1510.05135},
 primaryClass = {astro-ph.GA},
       adsurl = {https://ui.adsabs.harvard.edu/abs/2015A&A...584A.124F}
}

@ARTICLE{Furuya_2019,
       author = {{Furuya}, Kenji and {Aikawa}, Yuri and {Hama}, Tetsuya and {Watanabe}, Naoki},
        title = "{H$_{2}$ Ortho-Para Spin Conversion on Inhomogeneous Grain Surfaces}",
      journal = {\apj},
         year = 2019,
        month = sep,
       volume = {882},
       number = {2},
          eid = {172},
        pages = {172},
          doi = {10.3847/1538-4357/ab3790},
archivePrefix = {arXiv},
       eprint = {1908.01966},
 primaryClass = {astro-ph.GA},
       adsurl = {https://ui.adsabs.harvard.edu/abs/2019ApJ...882..172F}
}

@ARTICLE{Watanabe_2010,
       author = {{Watanabe}, Naoki and {Kimura}, Yuki and {Kouchi}, Akira and {Chigai}, Takeshi and {Hama}, Tetsuya and {Pirronello}, Valerio},
        title = "{Direct Measurements of Hydrogen Atom Diffusion and the Spin Temperature of Nascent H$_{2}$ Molecule on Amorphous Solid Water}",
      journal = {\apjl},
         year = 2010,
        month = may,
       volume = {714},
       number = {2},
        pages = {L233-L237},
          doi = {10.1088/2041-8205/714/2/L233},
       adsurl = {https://ui.adsabs.harvard.edu/abs/2010ApJ...714L.233W}
}

@ARTICLE{Walmsley_2004,
       author = {{Walmsley}, C.~M. and {Flower}, D.~R. and {Pineau des For{\^e}ts}, G.},
        title = "{Complete depletion in prestellar cores}",
      journal = {\aap},
         year = 2004,
        month = may,
       volume = {418},
        pages = {1035-1043},
          doi = {10.1051/0004-6361:20035718},
archivePrefix = {arXiv},
       eprint = {astro-ph/0402493},
 primaryClass = {astro-ph},
       adsurl = {https://ui.adsabs.harvard.edu/abs/2004A&A...418.1035W}
}

@ARTICLE{Flower_2006,
       author = {{Flower}, D.~R. and {Pineau Des For{\^e}ts}, G. and {Walmsley}, C.~M.},
        title = "{The importance of the ortho:para H$_{2}$ ratio for the deuteration of molecules during pre-protostellar collapse}",
      journal = {\aap},
         year = 2006,
        month = apr,
       volume = {449},
       number = {2},
        pages = {621-629},
          doi = {10.1051/0004-6361:20054246},
archivePrefix = {arXiv},
       eprint = {astro-ph/0601429},
 primaryClass = {astro-ph},
       adsurl = {https://ui.adsabs.harvard.edu/abs/2006A&A...449..621F}
}

@ARTICLE{Aikawa_2018,
       author = {{Aikawa}, Yuri and {Furuya}, Kenji and {Hincelin}, Ugo and {Herbst}, Eric},
        title = "{Multiple Paths of Deuterium Fractionation in Protoplanetary Disks}",
      journal = {\apj},
         year = 2018,
        month = mar,
       volume = {855},
       number = {2},
          eid = {119},
        pages = {119},
          doi = {10.3847/1538-4357/aaad6c},
archivePrefix = {arXiv},
       eprint = {1803.02498},
 primaryClass = {astro-ph.SR},
       adsurl = {https://ui.adsabs.harvard.edu/abs/2018ApJ...855..119A}
}

@ARTICLE{Tsuge_2019,
       author = {{Tsuge}, Masashi and {Hama}, Tetsuya and {Kimura}, Yuki and {Kouchi}, Akira and {Watanabe}, Naoki},
        title = "{Interactions of Atomic and Molecular Hydrogen with a Diamond-like Carbon Surface: H$_{2}$ Formation and Desorption}",
      journal = {\apj},
         year = 2019,
        month = jun,
       volume = {878},
       number = {1},
          eid = {23},
        pages = {23},
          doi = {10.3847/1538-4357/ab1e4e},
archivePrefix = {arXiv},
       eprint = {1911.05300},
 primaryClass = {astro-ph.GA},
       adsurl = {https://ui.adsabs.harvard.edu/abs/2019ApJ...878...23T}
}

@ARTICLE{Furuya_2026,
       author = {{Furuya}, Kenji and {Sugimoto}, Toshiki and {Iwasaki}, Kazunari and {Tsuge}, Masashi and {Watanabe}, Naoki},
        title = "{H$_{2}$ Ortho─Para Spin Conversion on Inhomogeneous Grain Surfaces. II. Impact of the Rotational Energy Difference between Adsorbed Ortho-H$_{2}$ and Para-H$_{2}$ and Implication for Deuterium Fractionation Chemistry}",
      journal = {\apj},
         year = 2026,
        month = apr,
       volume = {1001},
       number = {1},
          eid = {62},
        pages = {62},
          doi = {10.3847/1538-4357/ae43e3},
archivePrefix = {arXiv},
       eprint = {2602.13122},
 primaryClass = {astro-ph.GA},
       adsurl = {https://ui.adsabs.harvard.edu/abs/2026ApJ..1001...62F}
}

@ARTICLE{Hasegawa_1993b,
       author = {{Hasegawa}, T.~I. and {Herbst}, E.},
        title = "{Three-Phase Chemical Models of Dense Interstellar Clouds - Gas Dust Particle Mantles and Dust Particle Surfaces}",
      journal = {\mnras},
         year = 1993,
        month = aug,
       volume = {263},
        pages = {589},
          doi = {10.1093/mnras/263.3.589},
       adsurl = {https://ui.adsabs.harvard.edu/abs/1993MNRAS.263..589H}
}

@ARTICLE{Hily-Blant_2018,
       author = {{Hily-Blant}, P. and {Faure}, A. and {Rist}, C. and {Pineau des For{\^e}ts}, G. and {Flower}, D.~R.},
        title = "{Modelling the molecular composition and nuclear-spin chemistryof collapsing pre-stellar sources}",
      journal = {\mnras},
         year = 2018,
        month = jul,
       volume = {477},
       number = {4},
        pages = {4454-4472},
          doi = {10.1093/mnras/sty881},
archivePrefix = {arXiv},
       eprint = {1804.01354},
 primaryClass = {astro-ph.GA},
       adsurl = {https://ui.adsabs.harvard.edu/abs/2018MNRAS.477.4454H}
}

@ARTICLE{Honvault_2011,
       author = {{Honvault}, P. and {Jorfi}, M. and {Gonz{\'a}lez-Lezana}, T. and {Faure}, A. and {Pagani}, L.},
        title = "{Ortho-Para H$_{2}$ Conversion by Proton Exchange at Low Temperature: An Accurate Quantum Mechanical Study}",
      journal = {\prl},
         year = 2011,
        month = jul,
       volume = {107},
       number = {2},
          eid = {023201},
        pages = {023201},
          doi = {10.1103/PhysRevLett.107.023201},
       adsurl = {https://ui.adsabs.harvard.edu/abs/2011PhRvL.107b3201H}
}

@ARTICLE{Honvault_2012,
       author = {{Honvault}, P. and {Jorfi}, M. and {Gonz{\'a}lez-Lezana}, T. and {Faure}, A. and {Pagani}, L.},
        title = "{Erratum: Otho-Para H$_{2}$ Conversion by Proton Exchange at Low Temperature: An Accurate Quantum Mechanical Study [Phys. Rev. Lett. 107, 023201 (2011)]}",
      journal = {\prl},
         year = 2012,
        month = mar,
       volume = {108},
       number = {10},
          eid = {109903},
        pages = {109903},
          doi = {10.1103/PhysRevLett.108.109903},
       adsurl = {https://ui.adsabs.harvard.edu/abs/2012PhRvL.108j9903H}
}

@ARTICLE{Komichi_2024,
       author = {{Komichi}, Yuto and {Aikawa}, Yuri and {Iwasaki}, Kazunari and {Furuya}, Kenji},
        title = "{Chemical evolution during molecular cloud formation triggered by an interstellar shock wave: dependence on shock parameters and comparison with molecular absorption lines}",
      journal = {\mnras},
         year = 2024,
        month = dec,
       volume = {535},
       number = {4},
        pages = {3738-3757},
          doi = {10.1093/mnras/stae2599},
archivePrefix = {arXiv},
       eprint = {2411.12394},
 primaryClass = {astro-ph.GA},
       adsurl = {https://ui.adsabs.harvard.edu/abs/2024MNRAS.535.3738K}
}

@ARTICLE{Aikawa_1999,
       author = {{Aikawa}, Yuri and {Herbst}, Eric},
        title = "{Deuterium Fractionation in Protoplanetary Disks}",
      journal = {\apj},
         year = 1999,
        month = nov,
       volume = {526},
       number = {1},
        pages = {314-326},
          doi = {10.1086/307973},
       adsurl = {https://ui.adsabs.harvard.edu/abs/1999ApJ...526..314A}
}

@ARTICLE{Draine_1978,
       author = {{Draine}, B.~T.},
        title = "{Photoelectric heating of interstellar gas.}",
      journal = {\apjs},
         year = 1978,
        month = apr,
       volume = {36},
        pages = {595-619},
          doi = {10.1086/190513},
       adsurl = {https://ui.adsabs.harvard.edu/abs/1978ApJS...36..595D}
}

@ARTICLE{Brunken_2014,
       author = {{Br{\"u}nken}, Sandra and {Sipil{\"a}}, Olli and {Chambers}, Edward T. and {Harju}, Jorma and {Caselli}, Paola and {Asvany}, Oskar and {Honingh}, Cornelia E. and {Kami{\'n}ski}, Tomasz and {Menten}, Karl M. and {Stutzki}, J{\"u}rgen and et al.},
        title = "{H$_{2}$D$^{+}$ observations give an age of at least one million years for a cloud core forming Sun-like stars}",
      journal = {\nat},
         year = 2014,
        month = dec,
       volume = {516},
       number = {7530},
        pages = {219-221},
          doi = {10.1038/nature13924},
       adsurl = {https://ui.adsabs.harvard.edu/abs/2014Natur.516..219B}
}

@ARTICLE{Sternberg_1999,
       author = {{Sternberg}, Amiel and {Neufeld}, David A.},
        title = "{The Ratio of Ortho- to Para-H$_{2}$ in Photodissociation Regions}",
      journal = {\apj},
         year = 1999,
        month = may,
       volume = {516},
       number = {1},
        pages = {371-380},
          doi = {10.1086/307115},
archivePrefix = {arXiv},
       eprint = {astro-ph/9812049},
 primaryClass = {astro-ph},
       adsurl = {https://ui.adsabs.harvard.edu/abs/1999ApJ...516..371S}
}

@ARTICLE{Zannese_2025,
       author = {{Zannese}, M. and {Guillard}, P. and {Abergel}, A. and {Habart}, E. and {Dell'Ova}, P. and {Trahin}, B. and {Le Bourlot}, J. and {Misselt}, K. and {Van De Putte}, D. and {Witt}, A.~N. and {Gordon}, K.~D. and {Noriega-Crespo}, A. and {Baes}, M. and {Bouchet}, P. and {Brandl}, B.~R. and {Elyajouri}, M. and {Kannavou}, O. and {Klassen}, P. and {Ysard}, N.},
        title = "{JWST observations of photodissociation regions: II. Warm molecular hydrogen spectroscopy in the Horsehead nebula}",
      journal = {\aap},
         year = 2025,
        month = dec,
       volume = {704},
          eid = {A202},
        pages = {A202},
          doi = {10.1051/0004-6361/202556314},
archivePrefix = {arXiv},
       eprint = {2510.06812},
 primaryClass = {astro-ph.GA},
       adsurl = {https://ui.adsabs.harvard.edu/abs/2025A&A...704A.202Z}
}

@ARTICLE{Ichimura_2026,
       author = {{Ichimura}, Ryota and {Nomura}, Hideko and {Furuya}, Kenji and {Hama}, Tetsuya and {Millar}, T.~J.},
        title = "{Isotopomer-Specific Carbon Isotope Ratio of Complex Organic Molecules in Star-Forming Cores}",
      journal = {ACS Earth and Space Chemistry},
         year = 2026,
        month = jan,
       volume = {10},
       number = {1},
        pages = {43-56},
          doi = {10.1021/acsearthspacechem.5c00180},
archivePrefix = {arXiv},
       eprint = {2512.10516},
 primaryClass = {astro-ph.GA},
       adsurl = {https://ui.adsabs.harvard.edu/abs/2026ESC....10...43I}
}

@ARTICLE{Majumdar_2017,
       author = {{Majumdar}, L. and {Gratier}, P. and {Ruaud}, M. and {Wakelam}, V. and {Vastel}, C. and {Sipil{\"a}}, O. and {Hersant}, F. and {Dutrey}, A. and {Guilloteau}, S.},
        title = "{Chemistry of TMC-1 with multiply deuterated species and spin chemistry of H$_{2}$, H$_{2}$$^{+}$, H$_{3}$$^{+}$ and their isotopologues}",
      journal = {\mnras},
         year = 2017,
        month = apr,
       volume = {466},
       number = {4},
        pages = {4470-4479},
          doi = {10.1093/mnras/stw3360},
archivePrefix = {arXiv},
       eprint = {1612.07845},
 primaryClass = {astro-ph.GA},
       adsurl = {https://ui.adsabs.harvard.edu/abs/2017MNRAS.466.4470M}
}

@ARTICLE{Sipila_2017,
       author = {{Sipil{\"a}}, O. and {Harju}, J. and {Caselli}, P.},
        title = "{Species-to-species rate coefficients for the H$_{3}$$^{+}$ + H$_{2}$ reacting system}",
      journal = {\aap},
         year = 2017,
        month = oct,
       volume = {607},
          eid = {A26},
        pages = {A26},
          doi = {10.1051/0004-6361/201731039},
archivePrefix = {arXiv},
       eprint = {1707.03170},
 primaryClass = {astro-ph.GA},
       adsurl = {https://ui.adsabs.harvard.edu/abs/2017A&A...607A..26S}
}

@ARTICLE{Maret_2007,
       author = {{Maret}, S{\'e}bastien and {Bergin}, Edwin A.},
        title = "{The Ionization Fraction of Barnard 68: Implications for Star and Planet Formation}",
      journal = {\apj},
         year = 2007,
        month = aug,
       volume = {664},
       number = {2},
        pages = {956-963},
          doi = {10.1086/519152},
archivePrefix = {arXiv},
       eprint = {0704.3188},
 primaryClass = {astro-ph},
       adsurl = {https://ui.adsabs.harvard.edu/abs/2007ApJ...664..956M}
}

@ARTICLE{Pagani_2009,
       author = {{Pagani}, L. and {Vastel}, C. and {Hugo}, E. and {Kokoouline}, V. and {Greene}, C.~H. and {Bacmann}, A. and {Bayet}, E. and {Ceccarelli}, C. and {Peng}, R. and {Schlemmer}, S.},
        title = "{Chemical modeling of <ASTROBJ>L183</ASTROBJ> (<ASTROBJ>L134N</ASTROBJ>): an estimate of the ortho/para H\{\_2\} ratio}",
      journal = {\aap},
         year = 2009,
        month = feb,
       volume = {494},
       number = {2},
        pages = {623-636},
          doi = {10.1051/0004-6361:200810587},
archivePrefix = {arXiv},
       eprint = {0810.1861},
 primaryClass = {astro-ph},
       adsurl = {https://ui.adsabs.harvard.edu/abs/2009A&A...494..623P}
}

@ARTICLE{Pagani_2013,
       author = {{Pagani}, L. and {Lesaffre}, P. and {Jorfi}, M. and {Honvault}, P. and {Gonz{\'a}lez-Lezana}, T. and {Faure}, A.},
        title = "{Ortho-H$_{2}$ and the age of prestellar cores}",
      journal = {\aap},
         year = 2013,
        month = mar,
       volume = {551},
          eid = {A38},
        pages = {A38},
          doi = {10.1051/0004-6361/201117161},
       adsurl = {https://ui.adsabs.harvard.edu/abs/2013A&A...551A..38P}
}

@ARTICLE{Pagani_1992,
       author = {{Pagani}, L. and {Salez}, M. and {Wannier}, P.~G.},
        title = "{The chemistry of H2D+ in cold clouds.}",
      journal = {\aap},
         year = 1992,
        month = may,
       volume = {258},
        pages = {479-488},
       adsurl = {https://ui.adsabs.harvard.edu/abs/1992A&A...258..479P}
}

@ARTICLE{Cleeves_2014,
       author = {{Cleeves}, L. Ilsedore and {Bergin}, Edwin A. and {Adams}, Fred C.},
        title = "{Exclusion of Cosmic Rays in Protoplanetary Disks. II. Chemical Gradients and Observational Signatures}",
      journal = {\apj},
         year = 2014,
        month = oct,
       volume = {794},
       number = {2},
          eid = {123},
        pages = {123},
          doi = {10.1088/0004-637X/794/2/123},
archivePrefix = {arXiv},
       eprint = {1408.2835},
 primaryClass = {astro-ph.SR},
       adsurl = {https://ui.adsabs.harvard.edu/abs/2014ApJ...794..123C}
}

@ARTICLE{Lee_2015,
       author = {{Lee}, Jeong-Eun and {Bergin}, Edwin A.},
        title = "{The D/H Ratio of Water Ice at Low Temperatures}",
      journal = {\apj},
         year = 2015,
        month = jan,
       volume = {799},
       number = {1},
          eid = {104},
        pages = {104},
          doi = {10.1088/0004-637X/799/1/104},
archivePrefix = {arXiv},
       eprint = {1411.4231},
 primaryClass = {astro-ph.GA},
       adsurl = {https://ui.adsabs.harvard.edu/abs/2015ApJ...799..104L}
}

@ARTICLE{Fukutani_2013,
       author = {{Fukutani}, K. and {Sugimoto}, T.},
        title = "{Physisorption and ortho-para conversion of molecular hydrogen on solid surfaces}",
      journal = {Progress In Surface Science},
         year = 2013,
        month = dec,
       volume = {88},
       number = {4},
        pages = {279-348},
          doi = {10.1016/j.progsurf.2013.09.001},
       adsurl = {https://ui.adsabs.harvard.edu/abs/2013PrSS...88..279F}
}

@ARTICLE{Gerlich_1990,
       author = {{Gerlich}, D.},
        title = "{Ortho-para transitions in reactive H$^{ + }$+H$_{2}$ collisions}",
      journal = {\jcp},
         year = 1990,
        month = feb,
       volume = {92},
       number = {4},
        pages = {2377-2388},
          doi = {10.1063/1.457980},
       adsurl = {https://ui.adsabs.harvard.edu/abs/1990JChPh..92.2377G}
}

@ARTICLE{Tsuge_2021,
       author = {{Tsuge}, M. and {Namiyoshi}, T. and {Furuya}, K. and {Yamazaki}, T. and {Kouchi}, A. and {Watanabe}, N.},
        title = "{Rapid Ortho-to-para Nuclear Spin Conversion of H$_{2}$ on a Silicate Dust Surface}",
      journal = {\apj},
         year = 2021,
        month = feb,
       volume = {908},
       number = {2},
          eid = {234},
        pages = {234},
          doi = {10.3847/1538-4357/abd9c0},
archivePrefix = {arXiv},
       eprint = {2101.02357},
 primaryClass = {astro-ph.GA},
       adsurl = {https://ui.adsabs.harvard.edu/abs/2021ApJ...908..234T}
}

@ARTICLE{Hugo_2009,
       author = {{Hugo}, Edouard and {Asvany}, Oskar and {Schlemmer}, Stephan},
        title = "{H$_{3}$$^{+}$+H$_{2}$ isotopic system at low temperatures: Microcanonical model and experimental study}",
      journal = {\jcp},
         year = 2009,
        month = apr,
       volume = {130},
       number = {16},
        pages = {164302-164302},
          doi = {10.1063/1.3089422},
       adsurl = {https://ui.adsabs.harvard.edu/abs/2009JChPh.130p4302H}
}

@ARTICLE{Oka_2004,
       author = {{Oka}, Takeshi},
        title = "{Nuclear spin selection rules in chemical reactions by angular momentum algebra}",
      journal = {Journal of Molecular Spectroscopy},
         year = 2004,
        month = dec,
       volume = {228},
       number = {2},
        pages = {635-639},
          doi = {10.1016/j.jms.2004.08.015},
       adsurl = {https://ui.adsabs.harvard.edu/abs/2004JMoSp.228..635O}
}

@ARTICLE{Bovino_2021,
       author = {{Bovino}, Stefano and {Lupi}, Alessandro and {Giannetti}, Andrea and {Sabatini}, Giovanni and {Schleicher}, Dominik R.~G. and {Wyrowski}, Friedrich and {Menten}, Karl M.},
        title = "{Chemical analysis of prestellar cores in Ophiuchus yields short timescales and rapid collapse}",
      journal = {\aap},
         year = 2021,
        month = oct,
       volume = {654},
          eid = {A34},
        pages = {A34},
          doi = {10.1051/0004-6361/202141252},
archivePrefix = {arXiv},
       eprint = {2105.02253},
 primaryClass = {astro-ph.GA},
       adsurl = {https://ui.adsabs.harvard.edu/abs/2021A&A...654A..34B}
}

@ARTICLE{Lupi_2021,
       author = {{Lupi}, Alessandro and {Bovino}, Stefano and {Grassi}, Tommaso},
        title = "{On the low ortho-to-para H$_{2}$ ratio in star-forming filaments}",
      journal = {\aap},
         year = 2021,
        month = oct,
       volume = {654},
          eid = {L6},
        pages = {L6},
          doi = {10.1051/0004-6361/202142145},
archivePrefix = {arXiv},
       eprint = {2109.02655},
 primaryClass = {astro-ph.GA},
       adsurl = {https://ui.adsabs.harvard.edu/abs/2021A&A...654L...6L}
}

@ARTICLE{Bovion_2019,
       author = {{Bovino}, S. and {Ferrada-Chamorro}, S. and {Lupi}, A. and {Sabatini}, G. and {Giannetti}, A. and {Schleicher}, D.~R.~G.},
        title = "{The 3D Structure of CO Depletion in High-mass Prestellar Regions}",
      journal = {\apj},
         year = 2019,
        month = dec,
       volume = {887},
       number = {2},
          eid = {224},
        pages = {224},
          doi = {10.3847/1538-4357/ab53e4},
archivePrefix = {arXiv},
       eprint = {1910.13981},
 primaryClass = {astro-ph.GA},
       adsurl = {https://ui.adsabs.harvard.edu/abs/2019ApJ...887..224B}
}

@ARTICLE{Sipila_2013,
       author = {{Sipil{\"a}}, O. and {Caselli}, P. and {Harju}, J.},
        title = "{HD depletion in starless cores}",
      journal = {\aap},
         year = 2013,
        month = jun,
       volume = {554},
          eid = {A92},
        pages = {A92},
          doi = {10.1051/0004-6361/201220922},
archivePrefix = {arXiv},
       eprint = {1304.4031},
 primaryClass = {astro-ph.GA},
       adsurl = {https://ui.adsabs.harvard.edu/abs/2013A&A...554A..92S}
}

\begin{appendix}




\onecolumn
\section{Chemical reactions considered in equations for H$_3^+$, H$_2$D$^+$, D$_2$H$^+$, and D$_2$ OPR, and D$_3^+$ OMR}

Table \ref{table:OPRreaction_H3p} to \ref{table:OPRreaction_D2} summarize the chemical reactions considered in equations of H$_3^+$ OPR, H$_2$D$^+$ OPR, D$_2$H$^+$ OPR, D$_3^+$ OMR, and D$_2$ OPR, respectively.

\begin{table}[h!]
\caption{Similar to Table \ref{table:OPRreaction}, but for H$_3^+$.}                 
\label{table:OPRreaction_H3p}  
\centering                      
\begin{tabular}{l|l}     
\hline
\multicolumn{2}{c}{H$_3^+$ OPR (depending on H$_2$ OPR)} \\        
\hline
    Formation   & H$_2^+$ + o/p-H$_2$ $\rightarrow$ o/p-H$_3^+$ + H \\
\hline                  
    Exchange    & o-H$_3^+$ + HD $\rightarrow$ p-H$_3^+$ + HD \\
                & p-H$_3^+$ + HD $\rightarrow$ o-H$_3^+$ + HD \\
                & o-H$_3^+$ + o-H$_2$ $\rightarrow$ p-H$_3^+$ + o/p-H$_2$ \\
                & o-H$_3^+$ + p-H$_2$ $\rightarrow$ p-H$_3^+$ + o/p-H$_2$ \\
                & p-H$_3^+$ + o-H$_2$ $\rightarrow$ o-H$_3^+$ + o/p-H$_2$ \\         
                & p-H$_3^+$ + p-H$_2$ $\rightarrow$ o-H$_3^+$ + o/p-H$_2$ \\
\hline                    
    Destruction & o-H$_3^+$ + HD $\rightarrow$ o/p-H$_2$D$^+$ + o/p-H$_2$ \\
                & p-H$_3^+$ + HD $\rightarrow$ o/p-H$_2$D$^+$ + o/p-H$_2$ \\
                & o-H$_3^+$ + e$^-$ $\rightarrow$ o/p-H$_2$ + H, H + H + H \\
                & p-H$_3^+$ + e$^-$ $\rightarrow$ o/p-H$_2$ + H, H + H + H \\
\hline        
\end{tabular}
\end{table}

\begin{table}[h!]
\caption{Similar to Table \ref{table:OPRreaction}, but for H$_2$D$^+$.}                
\label{table:OPRreaction_H2Dp}   
\centering                     
\begin{tabular}{l|l}    
\hline                  
\multicolumn{2}{c}{H$_2$D$^+$ OPR (depending on H$_2$ and H$_3^+$ OPR)} \\       
\hline
    Formation   & o-H$_3^+$ + HD $\rightarrow$ o-H$_2$D$^+$ + o/p-H$_2$ \\
                & o-H$_3^+$ + HD $\rightarrow$ p-H$_2$D$^+$ + o/p-H$_2$ \\
                & p-H$_3^+$ + HD $\rightarrow$ o-H$_2$D$^+$ + o/p-H$_2$ \\
                & p-H$_3^+$ + HD $\rightarrow$ p-H$_2$D$^+$ + o/p-H$_2$ \\
\hline                     
    Exchange    & o-H$_2$D$^+$ + o-H$_2$ $\rightarrow$ p-H$_2$D$^+$ + o/p-H$_2$ \\
                & o-H$_2$D$^+$ + p-H$_2$ $\rightarrow$ p-H$_2$D$^+$ + o/p-H$_2$ \\
                & p-H$_2$D$^+$ + o-H$_2$ $\rightarrow$ o-H$_2$D$^+$ + o/p-H$_2$ \\
                & p-H$_2$D$^+$ + p-H$_2$ $\rightarrow$ o-H$_2$D$^+$ + o/p-H$_2$ \\
\hline                   
    Destruction & o-H$_2$D$^+$ + o-H$_2$ $\rightarrow$ o/p-H$_3^+$ + HD \\
                & o-H$_2$D$^+$ + p-H$_2$ $\rightarrow$ o/p-H$_3^+$ + HD \\
                & p-H$_2$D$^+$ + o-H$_2$ $\rightarrow$ o/p-H$_3^+$ + HD \\
                & p-H$_2$D$^+$ + p-H$_2$ $\rightarrow$ o/p-H$_3^+$ + HD \\
\hline       
\end{tabular}
\end{table}

\begin{table}[h!]
\caption{Similar to Table \ref{table:OPRreaction}, but for D$_2$H$^+$.}                
\label{table:OPRreaction_D2Hp}  
\centering                     
\begin{tabular}{l|l}     
\hline                      
\multicolumn{2}{c}{D$_2$H$^+$ OPR (depending on H$_2$, D$_2$, H$_3^+$ and H$_2$D$^+$ OPR)} \\     
\hline
    Formation   & o-H$_2$D$^+$ + HD $\rightarrow$ o-D$_2$H$^+$ + o/p-H$_2$ \\
                & o-H$_2$D$^+$ + HD $\rightarrow$ p-D$_2$H$^+$ + o/p-H$_2$ \\
                & p-H$_2$D$^+$ + HD $\rightarrow$ o-D$_2$H$^+$ + o/p-H$_2$ \\
                & p-H$_2$D$^+$ + HD $\rightarrow$ p-D$_2$H$^+$ + o/p-H$_2$ \\
                & o-H$_3^+$ + D$_2$ $\rightarrow$ o-D$_2$H$^+$ + o/p-H$_2$ \\
                & o-H$_3^+$ + D$_2$ $\rightarrow$ p-D$_2$H$^+$ + o/p-H$_2$ \\
                & p-H$_3^+$ + D$_2$ $\rightarrow$ o-D$_2$H$^+$ + o/p-H$_2$ \\
                & p-H$_3^+$ + D$_2$ $\rightarrow$ p-D$_2$H$^+$ + o/p-H$_2$ \\
\hline                  
    Exchange    & o-D$_2$H$^+$ + HD $\rightarrow$ p-D$_2$H$^+$ + HD \\
                & p-D$_2$H$^+$ + HD $\rightarrow$ o-D$_2$H$^+$ + HD \\
\hline                   
    Destruction & o-D$_2$H$^+$ + o-H$_2$ $\rightarrow$ o/p-H$_2$D$^+$ + HD \\
                & o-D$_2$H$^+$ + p-H$_2$ $\rightarrow$ o/p-H$_2$D$^+$ + HD \\
                & p-D$_2$H$^+$ + o-H$_2$ $\rightarrow$ o/p-H$_2$D$^+$ + HD \\
                & p-D$_2$H$^+$ + p-H$_2$ $\rightarrow$ o/p-H$_2$D$^+$ + HD \\
\hline                      
\end{tabular}
\end{table}

\begin{table}[h!]
\caption{Similar to Table\ref{table:OPRreaction_D3pOPR}, but for D$_3^+$ OMR}                
\label{table:OPRreaction_D3pOMR} 
\centering                     
\begin{tabular}{l|l}    
\hline                 
\multicolumn{2}{c}{D$_3^+$ OMR (depending on H$_2$, D$_2$, D$_2$H$^+$, and D$_3^+$ OPR)} \\    
\hline
    Formation   & o-D$_2$H$^+$ + HD $\rightarrow$ o-D$_3^+$ + HD \\
                & o-D$_2$H$^+$ + HD $\rightarrow$ m-D$_3^+$ + HD \\
                & p-D$_2$H$^+$ + HD $\rightarrow$ o-D$_3^+$ + HD \\
                & p-D$_2$H$^+$ + HD $\rightarrow$ m-D$_3^+$ + HD \\
                & p-D$_3^+$ + HD $\rightarrow$ o-D$_3^+$ + HD \\
                & p-D$_3^+$ + o-D$_2$ $\rightarrow$ o-D$_3^+$ + o/p-D$_2$ \\
                & p-D$_3^+$ + o-D$_2$ $\rightarrow$ m-D$_3^+$ + o/p-D$_2$ \\
                & p-D$_3^+$ + p-D$_2$ $\rightarrow$ o-D$_3^+$ + o/p-D$_2$ \\
                & p-D$_3^+$ + p-D$_2$ $\rightarrow$ m-D$_3^+$ + o/p-D$_2$ \\
\hline                   
    Exchange    & o-D$_3^+$ + HD $\rightarrow$ m-D$_3^+$ + HD \\
                & m-D$_3^+$ + HD $\rightarrow$ o-D$_3^+$ + HD \\
                & o-D$_3^+$ + o-D$_2$ $\rightarrow$ m-D$_3^+$ + o/p-D$_2$ \\
                & o-D$_3^+$ + p-D$_2$ $\rightarrow$ m-D$_3^+$ + o/p-D$_2$ \\
                & m-D$_3^+$ + o-D$_2$ $\rightarrow$ o-D$_3^+$ + o/p-D$_2$ \\
                & m-D$_3^+$ + p-D$_2$ $\rightarrow$ o-D$_3^+$ + o/p-D$_2$ \\
\hline                  
    Destruction & o-D$_3^+$ + o-H$_2$ $\rightarrow$ o/p-D$_2$H$^+$ + HD \\
                & o-D$_3^+$ + p-H$_2$ $\rightarrow$ o/p-D$_2$H$^+$ + HD \\
                & m-D$_3^+$ + o-H$_2$ $\rightarrow$ o/p-D$_2$H$^+$ + HD \\
                & m-D$_3^+$ + p-H$_2$ $\rightarrow$ o/p-D$_2$H$^+$ + HD \\
                & o-D$_3^+$ + HD $\rightarrow$ p-D$_3^+$ + HD \\
\hline
\end{tabular}
\end{table}

\begin{table}[h!]
\caption{Similar to Table\ref{table:OPRreaction}, but for D$_2$}                
\label{table:OPRreaction_D2}  
\centering                   
\begin{tabular}{l|l}     
\hline          
\multicolumn{2}{c}{D$_2$ OPR (depending on D$_2$H$^+$ and D$_3^+$ OPR, and D$_3^+$ OMR)} \\      
\hline
    Formation   & o-D$_2$H$^+$ + X $\rightarrow$ XH$^+$ + o-D$_2$ \\
                & p-D$_2$H$^+$ + X $\rightarrow$ XH$^+$ + p-D$_2$ \\
                & o-D$_3^+$ + X $\rightarrow$ XD$^+$ + o-D$_2$ \\
                & o-D$_3^+$ + X $\rightarrow$ XD$^+$ + p-D$_2$ \\
                & p-D$_3^+$ + X $\rightarrow$ XD$^+$ + p-D$_2$ \\
                & m-D$_3^+$ + X $\rightarrow$ XD$^+$ + o-D$_2$ \\
                & o-D$_2$H$^+$ + e$^-$ $\rightarrow$ o-D$_2$ + D \\
                & p-D$_2$H$^+$ + e$^-$ $\rightarrow$ p-D$_2$ + D \\
                & o-D$_3^+$ + e$^-$ $\rightarrow$ o-D$_2$ + D \\
                & o-D$_3^+$ + e$^-$ $\rightarrow$ p-D$_2$ + D \\
                & p-D$_3^+$ + e$^-$ $\rightarrow$ p-D$_2$ + D \\
                & m-D$_3^+$ + e$^-$ $\rightarrow$ m-D$_2$ + D \\
                & o-D$_2$H$^+$ + grain$^-$ $\rightarrow$ o-D$_2$ + D + grain$^0$\\
                & p-D$_2$H$^+$ + grain$^-$ $\rightarrow$ p-D$_2$ + D + grain$^0$ \\
                & o-D$_3^+$ + grain$^-$ $\rightarrow$ o-D$_2$ + D + grain$^0$ \\
                & o-D$_3^+$ + grain$^-$ $\rightarrow$ p-D$_2$ + D + grain$^0$ \\
                & p-D$_3^+$ + grain$^-$ $\rightarrow$ p-D$_2$ + D + grain$^0$ \\
                & m-D$_3^+$ + grain$^-$ $\rightarrow$ m-D$_2$ + D + grain$^0$ \\
\hline                    
    Exchange    & o-D$_3^+$ + p-D$_2$ $\rightarrow$ o/p/m-D$_3$ + o-D$_2$ \\
                & m-D$_3^+$ + p-D$_2$ $\rightarrow$ o/m-D$_3$ + o-D$_2$ \\
                & o-D$_3^+$ + o-D$_2$ $\rightarrow$ o/p/m-D$_3$ + p-D$_2$ \\
\hline                        
\end{tabular}
\tablefoot{X represents CO, O, N$_2$, OH, or H$_2$O. grain$^-$ and grain$^0$ represent negatively charged and neutral dust grains, respectively. The reactions between H$_2$ and H$_3$ isotopologues, which are not listed in the table, are also considered in both formation and destruction reactions. The references can be found in Section \ref{sec:setting}.}
\end{table}

\end{appendix}

\end{document}